\documentclass[11pt, a4paper]{article}

\usepackage[UKenglish]{babel}
\usepackage{amsmath, amssymb, hyperref, epsf}

\allowdisplaybreaks[4]

\hypersetup{
    colorlinks=true,
    linkcolor=black,
    citecolor=black,
    filecolor=black,
    urlcolor=black
}

\DeclareMathOperator{\im}{Im}
\DeclareMathOperator{\re}{Re}

\DeclareMathOperator{\const}{const}

\begin{document}

\pagenumbering{roman}

\begin{titlepage}

\begin{flushright}
{\bf \today} \\
DAMTP-11-32\\
MAD-TH-11-07\\

\end{flushright}

\begin{centering}

\vspace{0.5cm}

{ \Large{\bf Moduli Space and Wall-Crossing Formulae in\\ Higher-Rank Gauge Theories \par}}

\vspace{1cm}

Heng-Yu Chen${}^{1}$, Nick Dorey${}^{2}$ and Kirill Petunin${}^{2}$
\\[1cm]
${}^{1}$Department of Physics, University of Wisconsin,\\
Madison, WI 53706, USA
\\[0.5cm]
${}^{2}$DAMTP, Centre for Mathematical Sciences,\\
University of Cambridge, Wilberforce Road,\\
Cambridge, CB3 0WA, UK

\vspace{1cm}

{\bf \large Abstract} \\
\vspace{0.25cm}
We study the interplay between wall-crossing in four-dimensional gauge theory and instanton contributions to the moduli space metric of the same theory on $\mathbb{R}^{3}\times S^{1}$.
We consider $\mathcal{N}=2$ SUSY Yang--Mills with gauge group $SU(n)$ and focus on walls of marginal stability which extend to weak coupling.
By comparison with explicit field theory results we verify the Kontsevich--Soibelman formula for the change in the BPS spectrum at these walls and check the smoothness of the metric in the corresponding compactified theory.
We also verify in detail the predictions for the one instanton contribution to the metric coming from the non-linear integral equations of Gaiotto, Moore and Nietzke.
\end{centering}

\end{titlepage}

\pagenumbering{arabic}


\section{Introduction}

\paragraph{}

Gauge theories with $\mathcal{N}=2$ supersymmetry in four dimensions exhibit a rich variety of field-theoretic phenomena which can nevertheless be analysed precisely.
One area of recent progress is the spectrum of BPS states.
The index which counts these states is piecewise constant on the moduli space of vacua with discontinuities only at special submanifolds of codimension one known as walls of marginal stability.
Kontsevich and Soibelman \cite{KS} (KS) have conjectured an exact formula for the change in the spectrum as one of these walls is crossed.
Another related area of progress is in describing the Coulomb branch metric arising when the four-dimensional theory is compactified down to three dimensions on a circle.
In this context, the BPS states of the four-dimensional theory contribute as instantons after compactification.
A remarkable consequence of the KS wall-crossing formula, demonstrated in \cite{GMN}, is that it ensures the continuity of the Coulomb branch metric in the compactified theory and even determines the metric exactly via a set of non-linear integral equations.

\paragraph{}

As usual, it is instructive to compare proposed exact results in supersymmetric gauge theory with explicit semiclassical computations.
In a recent paper \cite{CDP} (see also \cite{CP}) we investigated the detailed structure of weak-coupling instanton corrections in $SU(2)$ gauge theory dictated by the integral equations of \cite{GMN}.
This resulted in a series of non-trivial predictions, which were then verified by comparison with systematic weak-coupling calculation.
The present paper is an extension of this investigation to the case of gauge group $SU(n)$.
For $n>2$ an important new phenomena arises: there are walls of marginal stability which extend to weak coupling.
This means that we can study the smoothness of the metric in the compactified theory, as predicted in \cite{GMN}, very explicitly.
We begin in section 2 by reviewing the weak-coupling spectrum of $\mathcal{N}=2$ SUSY Yang--Mills in four dimensions with gauge group $SU(n)$ and the corresponding walls of marginal stability.
We show that the Kontsevich--Soibelman wall-crossing formula agrees precisely with the jumps of the known semiclassical spectrum at these walls.
In section 3 we formulate the non-linear integral equations which are conjectured to determine the exact vacuum moduli space metric of the compactified theory.
We extract the leading weak-coupling corrections to the metric and verify the cancellation between single- and two-instanton contributions proposed in \cite{GMN} which renders the metric smooth.
In section 4 we compare the one-instanton contribution with the result of an explicit field theory calculation.
The main challenge here is to evaluate the non-cancelling ratio of functional determinants \cite{DKMTV} arising from fluctuations around the instanton background, which is a complicated function of the compactification radius.
Building on our earlier work \cite{CDP}, we express the resulting contribution in closed form and find a precise match with the prediction from the expansion of the integral equations described above.
Finally, after performing an appropriate Poisson resummation, we continue our results to zero radius and make contact with an earlier computation of Fraser and Tong \cite{Fraser Tong} in the three-dimensional theory.
The latter results also provide a direct check on the smoothness of the metric in the zero radius limit.


\section{Spectrum and wall-crossing in N=2 SU(n) gauge theories}

\paragraph{}

In this section, we shall first state the weak-coupling spectrum of the $\mathcal{N}=2$ $SU(n)$ theory, as derived in \cite{Fraser Hollowood} from the semiclassical monodromy of the associated Seiberg--Witten curve \cite{KLYT, AF}.
We then consider decay processes at the walls of marginal stability which extend into the weak-coupling region for $n>2$ and verify that Kontsevich--Soibelman wall-crossing formulae \cite{KS} precisely relate the spectra on both sides of each wall.
Combining these individual formulae, we find general equalities relating spectra in different weakly coupled regions of the moduli space. 


\subsection{Semiclassical spectrum}

\paragraph{}

We consider the $\mathcal{N}=2$ supersymmetric Yang--Mills theory \cite{SW} with gauge group $SU(n)$.
Let $r$ be the dimension of the Cartan subalgebra of the gauge group: for the $SU(n)$ group, $r=n-1$.
The gauge symmetry is maximally broken as $SU(n)\to U(1)^{n-1}$ by a vacuum expectation value (VEV) of the adjoint scalar field $\phi$.
$\langle\phi\rangle=\vec a \vec H$, where $\vec a$ is referred to as the ``electric coordinate'', $\vec H$ is the vector of matrices generating the Cartan subalgebra (all vectors are $r$-dimensional).
On the Coulomb branch of theory, the BPS spectrum consists of $n(n-1)/2$ massive pairs of $W^{\pm}$ bosons and tower of monopoles and dyons. 
Each BPS particle is labelled by its electric and magnetic charges under the residual $U(1)^{r-1}$ gauge groups:
\begin{equation}
\gamma = ({\vec \gamma}_{e},{\vec \gamma}_{m}) = \left( (\gamma_{e\,1},\dots,\gamma_{e\,r}),(\gamma_{m}^{1},\dots,\gamma_{m}^{r}) \right)
\,.
\end{equation}
Its central charge is given as
\begin{equation}
\label{central charge}
Z_{\gamma} = \vec a \vec\gamma_{e} + \vec a_{D} \vec\gamma_{m} = \sum_{I=1}^{r} \left( a^{I}\gamma_{e\,I}+a_{{D}\,I}\gamma_{m}^{I} \right)
\end{equation}
where $\vec a_{D}$ is the magnetic dual of $\vec a$, whose explicit form can be obtained from the holomorphic prepotential $\mathcal{F}(\vec a)$ \cite{KLYT,AF}, and the particle's mass is the modulus $|Z_{\gamma}|$.

\paragraph{}

Here we would like to discuss the semiclassical spectrum of the theory, following \cite{Fraser Hollowood}.
Denote the set of all roots of the gauge group $SU(n)$ as $\Phi$, the set of the $r=n-1$ simple roots as $\Phi_{0}$,
and the set of the $n(n-1)/2$ positive roots as $\Phi_{+}$.
Each $W^{+}$ boson corresponds to a positive root $\vec\alpha_{A}\in\Phi_{+}$ (and vice versa), so that it has charge $W_{A}=(\vec\alpha_{A},\vec 0)$ and mass $M_{W_{A}}=|\vec\alpha_{A}\vec a|$.
In the $SU(n)$ case, we will normalise every root $\vec\alpha$ as $\|\vec\alpha\|=1$.
By default, all roots will be denoted by Greek letters ($\vec\alpha$, \dots), positive roots will have capital Latin indices ($A$, \dots), and simple roots will have small Latin indices ($i$, \dots).
In terms of an orthonormal basis $\vec e_{i}$~\footnote{
This basis has $n=r+1$ dimensions, whereas all other vectors being considered are restricted to lie in $n-1=r$ dimensions.
}, simple roots for the $SU(n)$ group can be set as
\begin{equation}
\vec\alpha_{i} = \frac{1}{\sqrt{2}} \left( \vec e_{i}-\vec e_{i+1} \right)
\,,
\quad
i = \overline{1;n}
\,.
\end{equation}

\paragraph{}

For the $SU(3)$ gauge group, there exists a set of 3 positive roots, which may be chosen as $\Phi_{+}=\{\vec\alpha_{1}=(1,0),\ \vec\alpha_{2}=(-1/2,\sqrt{3}/2),\ \vec\alpha_{3}=(1/2,\sqrt{3}/2)\}$, where $\Phi_{0}=\{(1,0),(-1/2,\sqrt{3}/2)\}$ is a set of simple roots for the two-dimensional Cartan subalgebra.
For the $SU(2)$ gauge group, there is only one positive root, $1$, which is also simple.

\paragraph{}

We will be dealing with the weak-coupling region:
\begin{equation}
\left| \frac{\vec\alpha_{A}\vec a}{\Lambda} \right| \gg 1
\,,
\quad
\forall \ \vec\alpha_{A} \in \Phi_{+}
\end{equation}
where $\Lambda$ is the dynamical scale.
The global gauge transformations are not completely fixed: one can still perform discrete transformations in the Weyl group.
This discrete degree of freedom can be eliminated by requiring that $\re\vec a$ lies in the fundamental Weyl chamber corresponding to some choice of positive roots:
\begin{equation}
\label{Weyl chamber}
\re\left(\vec\alpha_{i}\vec a\right) \ge 0
\,,
\quad
\forall \ \vec\alpha_{i} \in \Phi_{0}
\,.
\end{equation}

\paragraph{}

The spectrum of dyons whose magnetic charge-vectors are given by simple roots (``simple dyons'') is analogous to the $SU(2)$ case:
\begin{equation}
\label{simple dyon}
\left( p\vec\alpha_{i},\vec\alpha_{i} \right)
\,,
\quad
\vec\alpha_{i} \in \Phi_{0}
\,, \
p \in \mathbb{Z}
\,.
\end{equation}

\paragraph{}

At weak coupling, to the leading order, the magnetic coordinate is given as
\footnote{
In \cite{Fraser Tong}, the convention is $\|\vec\alpha_{A}\|=2$, and therefore, the resulting coefficient is divided by 2.
}
\begin{equation}
\vec a_{D} = \frac{i}{\pi}\sum_{\vec\alpha_{A}\in\Phi_{+}} \vec\alpha_{A}(\vec\alpha_{A}\vec a) \,
\log \left( \frac{\vec\alpha_{A}\vec a}{\Lambda} \right)^{2} =
\hat\tau_{\rm eff} \vec a
\end{equation}
where $\hat\tau_{\rm eff}$ is the effective complex coupling.
The important feature of this expression is that it has singularities when one of the bosons becomes massless.
Following \cite{Fraser Hollowood}, for each singularity $\vec\alpha_{i}\vec a=0$, there should be a Weyl reflection \cite{Fraser Hollowood} acting on the VEV when $\re\vec\alpha_{i}\vec a=0$ to ensure that it stays within the fundamental Weyl chamber (\ref{Weyl chamber}).
This transformation reflects the projection of $\vec a$ onto $\vec\alpha_{i}$:

\begin{equation}
\vec a(t) = \vec a-\vec\alpha_{i} \left( \vec\alpha_{i}\vec a \right) \left( 1-e^{it} \right)
\,,
\quad
0 \le t \le \pi
\,,
\end{equation}
where $t=0$ and $t=\pi$ correspond to the initial and the final position (as $t$ increases, $\vec a(t)$ moves counterclockwise).
The associated monodromy matrix $\hat{M}_{i}$ acting on the vector $\left(\vec a,\vec a_{D}\right)$ from the left, and its inverse are given as
\begin{equation}
\hat{M}_{i} =
\left( 
\begin{array}{cc} 
\hat{1}-2\vec\alpha_{i}\otimes\vec\alpha_{i} & \hat{0} \\ 
-2\vec\alpha_{i}\otimes\vec\alpha_{i} & \hat{1}-2\vec\alpha_{i}\otimes\vec\alpha_{i}
\end{array}
\right)
\,,
\quad
\hat{M}_{i}^{-1} =
\left( 
\begin{array}{cc} 
\hat{1}-2\vec\alpha_{i}\otimes\vec\alpha_{i} & \hat{0} \\ 
2\vec\alpha_{i}\otimes\vec\alpha_{i} & \hat{1}-2\vec\alpha_{i}\otimes\vec\alpha_{i}
\end{array}
\right)
\,.
\end{equation}

\paragraph{}

We shall follow the approach in \cite{Fraser Hollowood} to obtain the full spectrum of dyons.
The dyons whose magnetic charges are not simple roots (``composite dyons'') are generated by acting on simple dyons with these monodromies (from the right) up to their overall sign.
The resulting spectrum of composite dyons is
\begin{equation}
\label{composite dyon}
\begin{aligned}
\left( p\vec\alpha_{i},\vec\alpha_{i} \right)
\hat{M}_{i+1}^{\epsilon_{i+1}}\hat{M}_{i+2}^{\epsilon_{i+2}}\dots\hat{M}_{j-1}^{\epsilon_{j-1}} & =
\left(
p\sum_{m=i}^{j-1}\vec\alpha_{m} +
\sum_{l=i+1}^{j-1}\epsilon_{l} \sum_{m=l}^{j-1}\vec\alpha_{m}
\,,\,
\sum_{m=i}^{j-1}\vec\alpha_{m}
\right)
\\
& = \frac{1}{\sqrt{2}} \left(
p \left( \vec e_{i}-\vec e_{j} \right) +
\sum_{l=i+1}^{j-1}\epsilon_{l} \left( \vec e_{l}-\vec e_{j} \right)
\,,\,
\vec e_{i}-\vec e_{j}
\right)
\,
\end{aligned}
\end{equation}
where $\epsilon_{l}=\pm 1$ (as for $|\epsilon_{l}|>1$, the VEV would cross a wall of marginal stability), $1\le j\le n$.
In the theory with gauge group $SU(3)$, (\ref{composite dyon}) has only one monodromy matrix, $\hat M_{2}$, and the composite dyons are $(p\vec\alpha_{3}\pm\vec\alpha_{2},\vec\alpha_{3})$, where $\vec\alpha_{3}=\vec\alpha_{1}+\vec\alpha_{2}$, depending on whether one acts with $\hat M_{2}$ or $\hat M_{2}^{-1}$ on $(p\vec\alpha_{1},\vec\alpha_{1})$.
This demonstrates that the moduli space of the $SU(3)$ theory at weak coupling consists of two separate regions.

\paragraph{}

Summing up, the spectrum is given by the sets of simple dyons (\ref{simple dyon}), composite dyons (\ref{composite dyon}), $W$ bosons with charge $(\vec\alpha_{A},\vec 0)$, and their antiparticles.


\subsection{Wall-crossing formulae}

\paragraph{}

As we have already mentioned, in $\mathcal{N}=2$ theories with gauge group $SU(n)$, $n\ge 3$, the weak-coupling spectrum is different in different regions of the moduli space.
These regions are separated by the so-called walls of marginal stability: on each wall, one composite dyon becomes unstable and decays (or, conversely, a bound state gets created).
Such decays are possible when the total central charge (\ref{central charge}) and the total mass are preserved.
For the decay process $\gamma\to\gamma_{1}+\gamma_{2}$, the conditions are $Z_{\gamma}=Z_{\gamma_{1}}+Z_{\gamma_{2}}$, $|Z_{\gamma}|=|Z_{\gamma_{1}}|+|Z_{\gamma_{2}}|$; this means that $\arg Z_{\gamma_{1}}=\arg Z_{\gamma_{2}}$.
To the leading order at weak coupling, the values of central charges for dyons depend only on their magnetic charges.
Hence, in this limit, the walls of marginal stability are given by
\begin{equation}
\label{weak wall}
\frac{\vec\alpha_{A}\vec a}{\vec\alpha_{B}\vec a} \in \mathbb{R}_{+}
\end{equation}
for some pair of positive roots, $\vec\alpha_{A}$ and $\vec\alpha_{B}$.
The composite dyon given by (\ref{composite dyon}) decays near the wall of marginal stability which can be reparametrised as
\begin{equation}
\frac{\sum_{m=i}^{k}\vec\alpha_{m}\vec a}{\sum_{m=k+1}^{j-1}\vec\alpha_{m}\vec a} \in \mathbb{R}_{+}
\quad \iff \quad
\frac{\vec e_{i}\vec a}{\vec e_{j}\vec a} \in \mathbb{R}_{+}
\,.
\end{equation}

\paragraph{}

When we take into account the electric charges of dyons corresponding to a given positive root, there is, in fact, no single wall of marginal stability, but rather, a collection of walls.
For every composite dyon, there is an individual wall where it can decay.
On the other hand, taking the effective coupling constant $g_{\rm eff}$ sufficiently small, all these individual walls can be approximated by (\ref{weak wall}); this is the reason why for the VEV far from (\ref{weak wall}), all such walls can be treated as a single wall given by (\ref{weak wall}).

\paragraph{}

Using the fact that each composite dyon can be parametrised as (\ref{composite dyon}), we can write down the decay processes:
\begin{equation}
\label{composite dyon decay}
\begin{aligned}
\pm \left(
p\sum_{m=i}^{j-1}\vec\alpha_{m}+\sum_{l=i+1}^{j-1}\epsilon_{l}\sum_{m=l}^{j-1}\vec\alpha_{m}
\,, \
\sum_{m=i}^{j-1}\vec\alpha_{m}
\right)
\to
\pm \left(
p\sum_{m=i}^{k}\vec\alpha_{m}+\sum_{l=i+1}^{k}\epsilon_{l}\sum_{m=l}^{k}\vec\alpha_{m}
\,, \
\sum_{m=i}^{k}\vec\alpha_{m}
\right)
\\
\pm \left(
\left( p+\sum_{l=i+1}^{k}\epsilon_{l} \right)\sum_{m=k+1}^{j-1}\vec\alpha_{m}+\sum_{l=k+1}^{j-1}\epsilon_{l}\sum_{m=l}^{j-1}\vec\alpha_{m}
\,, \
\sum_{m=k+1}^{j-1}\vec\alpha_{m}
\right)
\,,
\end{aligned}
\end{equation}
or, rewriting it in terms of the orthonormal basis introduced above,
\begin{equation}
\begin{aligned}
\pm \frac{1}{\sqrt{2}} & \left(
p\left( \vec e_{i}-\vec e_{j} \right)+\sum_{l=i+1}^{j-1}\epsilon_{l}\left( \vec e_{l}-\vec e_{j} \right)
\,, \
\vec e_{i}-\vec e_{j}
\right)
\\
\to
\pm \frac{1}{\sqrt{2}} & \left(
p\left( \vec e_{i}-\vec e_{k+1} \right)+\sum_{l=i+1}^{k}\epsilon_{l}\left( \vec e_{l}-\vec e_{k+1} \right)
\,, \
\vec e_{i}-\vec e_{k+1}
\right)
\\
\pm \frac{1}{\sqrt{2}} & \left(
\left( p+\sum_{l=i+1}^{k}\epsilon_{l} \right)\left( \vec e_{k+1}-\vec e_{j} \right)+\sum_{l=k+1}^{j-1}\epsilon_{l}\left( \vec e_{l}-\vec e_{j} \right)
\,, \
\vec e_{k+1}-\vec e_{j}
\right)
\,.
\end{aligned}
\end{equation}
In particular, for gauge group $SU(3)$, with one possible weak-coupling wall (\ref{weak wall}), $\vec\alpha_{1}\vec a/\vec\alpha_{2}\vec a\in\mathbb{R}_{+}$, there are two types of decays corresponding to the VEV approaching the wall from different sides:
\begin{equation}
\label{SU(3) composite dyon decay}
\begin{aligned}
\pm(p(\vec\alpha_{1}+\vec\alpha_{2})+\vec\alpha_{1},\vec\alpha_{1}+\vec\alpha_{2}) & \to
\pm((p+1)\vec\alpha_{1},\vec\alpha_{1})\pm(p\vec\alpha_{2},\vec\alpha_{2})
\,,
\\
\pm(p(\vec\alpha_{1}+\vec\alpha_{2})+\vec\alpha_{2},\vec\alpha_{1}+\vec\alpha_{2}) & \to
\pm((p+1)\vec\alpha_{2},\vec\alpha_{2})\pm(p\vec\alpha_{1},\vec\alpha_{1})
\,.
\end{aligned}
\end{equation}

\paragraph{}

Our goal is to express the spectra and decays discussed above in terms of Kontsevich--Soibelman operators \cite{KS} and show that the wall-crossing formulae are satisfied.
First, we need to introduce several definitions, following \cite{GMN}.
The symplectic product of two charges,
\begin{equation}
\gamma = \left( (\gamma_{e\,1},\dots,\gamma_{e\,r}),(\gamma_{m}^{1},\dots,\gamma_{m}^{r}) \right)
\,,
\quad
\xi = \left( (\xi_{e\,1},\dots, \xi_{e\,r}),(\xi_{m}^{1},\dots,\xi_{m}^{r}) \right)
\,,
\end{equation}
is defined as
\begin{equation}
\label{symplectic product}
\langle \gamma, \xi \rangle = -\vec\gamma_{e}\vec\xi_{m}+\vec\gamma_{m}\vec\xi_{e} = \sum_{I=1}^{r}(-\gamma_{e\,I} \xi_{m}^{I} + \gamma_{m}^{I} \xi_{e\,I})
\,.
\end{equation}
For each BPS particle with charge $\gamma$, associate the BPS ray $l_{\gamma}$ determined by the central charge $Z_{\gamma}(\vec a)$ (\ref{central charge}) in the complex plane (parametrised by an auxiliary variable $\zeta\in\mathbb{C}$):
\begin{equation}
\label{BPS ray}
l_{\gamma} = \left\{ \zeta \ : \ \frac{Z_{\gamma}(\vec a)}{\zeta} \in \mathbb{R}_{-} \right\}
\,.
\end{equation}
We also define a basis of Darboux coordinates (they will be used to find the moduli space metric): $\mathcal{X}_{e}^{I}$ and $\mathcal{X}_{m\,I}$ are called ``electric'' and ``magnetic'' components, and the index $I=1,\dots, r$.
More generally, Darboux coordinates for arbitrary electromagnetic charges are defined as
\begin{equation}
\label{Darboux coordinate}
\mathcal{X}_{\gamma}(\zeta) = \prod_{I=1}^{r} \left(\mathcal{X}_{e}^{I}(\zeta)\right)^{\gamma_{e\,I}} \left(\mathcal{X}_{m\,I}(\zeta)\right)^{\gamma_{m}^{I}}
\,,
\end{equation}
so that $\mathcal{X}_{e}^{I}$ and $\mathcal{X}_{m\,J}$ are the coordinates for charges with only one non-zero component, $\gamma_{e\,I}=1$ and $\gamma_{m}^{J}=1$, respectively.
Kontsevich--Soibelman operators are symplectomorphisms acting on Darboux coordinates as~\footnote{
Note that our conventions for the operators differ from \cite{KS, GMN} as we divided electric charge by two, so that electric charges of the pure theory are any integers (allowing half-integers in theories with flavours).
In \cite{KS, GMN}, $\mathcal{K}_{\gamma}$ acts as $\mathcal{X}_{\beta} \to \mathcal{X}_{\beta} \left(1-\sigma(\gamma)\mathcal{X}_{\gamma}\right)^{\langle\beta,\gamma\rangle}$ where $\sigma(\gamma)=(-1)^{\vec\gamma_{e}\vec\gamma_{m}}$.
}
\begin{equation}
\label{KS operator}
\mathcal{K}_{\gamma} \quad \colon \quad
\mathcal{X}_{\beta} \to \mathcal{X}_{\beta} \left( 1-\sigma(\gamma)\mathcal{X}_{\gamma} \right)^{2\langle\beta,\gamma\rangle}
\end{equation}
where $\sigma(\gamma)$ is the so-called quadratic refinement given by
\begin{equation}
\label{quadratic refinement}
\sigma(\gamma) = (-1)^{2\vec\gamma_{e}\vec\gamma_{m}} = (-1)^{2\sum_{I=1}^{r}\gamma_{e\,I}\gamma_{m}^{I}}
\,.
\end{equation}
We associate the following operator to each point $\vec a$ in the moduli space:
\begin{equation}
S = \prod_{\gamma\in\Gamma(\vec a)} \mathcal{K}_{\gamma}^{\Omega(\gamma,\vec a)}
\end{equation}
where $\Omega(\gamma,\vec a)$ is the degeneracy of the BPS state with charge $\gamma$ (giving $+1$ for dyons and $-2$ for $W$ bosons); all operators (i.e., their BPS rays) are ordered clockwise (equivalently, their central charges as complex vectors are ordered counterclockwise).
It was suggested in \cite{KS, GMN} that, although the spectrum changes across the moduli space, the resulting product $S$ is constant, and all such products are related by wall-crossing formulae.

\paragraph{}

We will verify this statement in the case of $SU(n)$ theories by constructing the formulae explicitly.
We will be using the following form of the pentagon wall-crossing formula for the decay processes under consideration:
\begin{equation}
\label{pentagon}
\mathcal{K}_{\gamma_{1}}\mathcal{K}_{\gamma_{2}} =
\mathcal{K}_{\gamma_{2}}\mathcal{K}_{\gamma_{1}+\gamma_{2}}\mathcal{K}_{\gamma_{1}}
\,,
\quad
\forall \ \langle\gamma_{1},\gamma_{2}\rangle=\pm\frac{1}{2}
\,.
\end{equation}
This is an extension of the rank 1 formula to higher rank gauge groups.
See appendix \ref{sec: pentagon} for the derivation of the formula and its applications to the $SU(n)$ case.

\paragraph{}

Let us start by considering the theory with gauge group $SU(3)$.
We notice that the pentagon formula describes decays of composite dyons (\ref{SU(3) composite dyon decay}):
\begin{equation}
\label{SU(3) pentagon}
\mathcal{K}_{\pm(p\vec\alpha_{1},\vec\alpha_{1})}
\mathcal{K}_{\pm((p+1)\vec\alpha_{2},\vec\alpha_{2})}
=
\mathcal{K}_{\pm((p+1)\vec\alpha_{2},\vec\alpha_{2})}
\mathcal{K}_{\pm(p(\vec\alpha_{2}+\vec\alpha_{1})+\vec\alpha_{2},\vec\alpha_{1}+\vec\alpha_{2})}
\mathcal{K}_{\pm(p\vec\alpha_{1},\vec\alpha_{1})}
\,.
\end{equation}
Starting with these formulae, we will construct the wall-crossing formula for the pure $SU(3)$ theory at weak coupling.
It is related to the wall-crossing formula \cite{KS} for the pure $SU(2)$ theory, which is given by
\begin{equation}
\label{WCF2}
\mathcal{K}_{(1,-1)}\mathcal{K}_{(0,1)}
=
\mathcal{K}_{(0,1)}\mathcal{K}_{(1,1)}\mathcal{K}_{(2,1)}\mathcal{K}_{(3,1)}
\dots \mathcal{K}_{(1,0)}^{-2} \dots
\mathcal{K}_{(4,-1)}\mathcal{K}_{(3,-1)}\mathcal{K}_{(2,-1)}\mathcal{K}_{(1,-1)}
\,.
\end{equation}
For any pair $\gamma$ and $-\gamma$ of particles from the spectrum, it contains only one operator, $\mathcal{K}_{\gamma}$ or $\mathcal{K}_{-\gamma}$.
To consider all particles preserving the order of operators, both sides of the formula should be multiplied (from the left or from the right) by the same expression, but with opposite charges.

\paragraph{}

In the $SU(n)$ case, we will require electric charge-vectors of $W$ bosons and magnetic charge-vectors of dyons to be positive roots, ignoring their antiparticles as they can be treated analogously.
Let us begin by writing out the wall-crossing formula implied by the known spectra of the $SU(3)$ theory on either side of the walls of marginal stability.
\begin{equation}
\label{WCF3}
\begin{aligned}
& \dots
\mathcal{K}_{(-2\vec\alpha_{1},\vec\alpha_{1})}
\mathcal{K}_{(-3(\vec\alpha_{1}+\vec\alpha_{2})+\vec\alpha_{1},\vec\alpha_{1}+\vec\alpha_{2})}
\mathcal{K}_{(-3\vec\alpha_{2},\vec\alpha_{2})}
\times
\mathcal{K}_{(-\vec\alpha_{1},\vec\alpha_{1})}
\mathcal{K}_{(-2(\vec\alpha_{1}+\vec\alpha_{2})+\vec\alpha_{1},\vec\alpha_{1}+\vec\alpha_{2})}
\mathcal{K}_{(-2\vec\alpha_{2},\vec\alpha_{2})}
\times \\ &
\mathcal{K}_{(\vec 0,\vec\alpha_{1})}
\mathcal{K}_{(-(\vec\alpha_{1}+\vec\alpha_{2})+\vec\alpha_{1},\vec\alpha_{1}+\vec\alpha_{2})}
\mathcal{K}_{(-\vec\alpha_{2},\vec\alpha_{2})}
\times
\mathcal{K}_{(\vec\alpha_{1},\vec\alpha_{1})}
\mathcal{K}_{(\vec\alpha_{1},\vec\alpha_{1}+\vec\alpha_{2})}
\mathcal{K}_{(\vec 0,\vec\alpha_{2})}
\times \\ &
\mathcal{K}_{(2\vec\alpha_{1},\vec\alpha_{1})}
\mathcal{K}_{((\vec\alpha_{1}+\vec\alpha_{2})+\vec\alpha_{1},\vec\alpha_{1}+\vec\alpha_{2})}
\mathcal{K}_{(\vec\alpha_{2},\vec\alpha_{2})}
\times
\mathcal{K}_{(3\vec\alpha_{1},\vec\alpha_{1})}
\mathcal{K}_{(2(\vec\alpha_{1}+\vec\alpha_{2})+\vec\alpha_{1},\vec\alpha_{1}+\vec\alpha_{2})}
\mathcal{K}_{(2\vec\alpha_{2},\vec\alpha_{2})}
\dots
\\
&
\mathcal{K}_{(\vec\alpha_{1},\vec 0)}^{-2}
\mathcal{K}_{(\vec\alpha_{1}+\vec\alpha_{2},\vec 0)}^{-2}
\mathcal{K}_{(\vec\alpha_{2},\vec 0)}^{-2}
\ =
\\
& \dots
\mathcal{K}_{(-2\vec\alpha_{2},\vec\alpha_{2})}
\mathcal{K}_{(-3(\vec\alpha_{1}+\vec\alpha_{2})+\vec\alpha_{2},\vec\alpha_{1}+\vec\alpha_{2})}
\mathcal{K}_{(-3\vec\alpha_{1},\vec\alpha_{1})}
\times
\mathcal{K}_{(-\vec\alpha_{2},\vec\alpha_{2})}
\mathcal{K}_{(-2(\vec\alpha_{1}+\vec\alpha_{2})+\vec\alpha_{2},\vec\alpha_{1}+\vec\alpha_{2})}
\mathcal{K}_{(-2\vec\alpha_{1},\vec\alpha_{1})}
\times \\ &
\mathcal{K}_{(\vec 0,\vec\alpha_{2})}
\mathcal{K}_{(-(\vec\alpha_{1}+\vec\alpha_{2})+\vec\alpha_{2},\vec\alpha_{1}+\vec\alpha_{2})}
\mathcal{K}_{(-\vec\alpha_{1},\vec\alpha_{1})}
\times
\mathcal{K}_{(\vec\alpha_{2},\vec\alpha_{2})}
\mathcal{K}_{(\vec\alpha_{2},\vec\alpha_{1}+\vec\alpha_{2})}
\mathcal{K}_{(\vec 0,\vec\alpha_{1})}
\times \\ &
\mathcal{K}_{(2\vec\alpha_{2},\vec\alpha_{2})}
\mathcal{K}_{((\vec\alpha_{1}+\vec\alpha_{2})+\vec\alpha_{2},\vec\alpha_{1}+\vec\alpha_{2})}
\mathcal{K}_{(\vec\alpha_{1},\vec\alpha_{1})}
\times
\mathcal{K}_{(3\vec\alpha_{2},\vec\alpha_{2})}
\mathcal{K}_{(2(\vec\alpha_{1}+\vec\alpha_{2})+\vec\alpha_{2},\vec\alpha_{1}+\vec\alpha_{2})}
\mathcal{K}_{(2\vec\alpha_{1},\vec\alpha_{1})}
\dots
\\
&
\mathcal{K}_{(\vec\alpha_{2},\vec 0)}^{-2}
\mathcal{K}_{(\vec\alpha_{1}+\vec\alpha_{2},\vec 0)}^{-2}
\mathcal{K}_{(\vec\alpha_{1},\vec 0)}^{-2}
\,,
\end{aligned}
\end{equation}
where all terms are multiplied in the group and the explicit notation ``$\times$'' is used only to group the terms in a convenient way.
The BPS ray at which the ordering starts is chosen differently from the $SU(2)$ case for further convenience.

\paragraph{}

We will now verify (\ref{WCF3}) by evaluating both sides.  
We can see how both sides of (\ref{WCF3}) change when the VEV passes thorough the walls.
For each decaying composite dyon, there is a corresponding pentagon identity (\ref{SU(3) pentagon}) modifying a fragment (separated by ``$\times$'') in (\ref{WCF3}).
Close to the wall (\ref{weak wall}), when all composite dyons decay, the wall-crossing formula (\ref{WCF3}) reduces to
\begin{equation}
\label{WCF3.1}
\begin{aligned}
& \dots
\mathcal{K}_{(-3\vec\alpha_{2},\vec\alpha_{2})}
\mathcal{K}_{(-2\vec\alpha_{1},\vec\alpha_{1})}
\times
\mathcal{K}_{(-2\vec\alpha_{2},\vec\alpha_{2})}
\mathcal{K}_{(-\vec\alpha_{1},\vec\alpha_{1})}
\times
\mathcal{K}_{(-\vec\alpha_{2},\vec\alpha_{2})}
\mathcal{K}_{(\vec 0,\vec\alpha_{1})}
\times \\ &
\mathcal{K}_{(\vec 0,\vec\alpha_{2})}
\mathcal{K}_{(\vec\alpha_{1},\vec\alpha_{1})}
\times
\mathcal{K}_{(\vec\alpha_{2},\vec\alpha_{2})}
\mathcal{K}_{(2\vec\alpha_{1},\vec\alpha_{1})}
\times
\mathcal{K}_{(2\vec\alpha_{2},\vec\alpha_{2})}
\mathcal{K}_{(3\vec\alpha_{1},\vec\alpha_{1})}
\dots \\ &
\mathcal{K}_{(\vec\alpha_{1},\vec 0)}^{-2}
\mathcal{K}_{(\vec\alpha_{1}+\vec\alpha_{2},\vec 0)}^{-2}
\mathcal{K}_{(\vec\alpha_{2},\vec 0)}^{-2}
\ =
\\
& \dots
\mathcal{K}_{(-3\vec\alpha_{1},\vec\alpha_{1})}
\mathcal{K}_{(-2\vec\alpha_{2},\vec\alpha_{2})}
\times
\mathcal{K}_{(-2\vec\alpha_{1},\vec\alpha_{1})}
\mathcal{K}_{(-\vec\alpha_{2},\vec\alpha_{2})}
\times
\mathcal{K}_{(-\vec\alpha_{1},\vec\alpha_{1})}
\mathcal{K}_{(\vec 0,\vec\alpha_{2})}
\times \\ &
\mathcal{K}_{(\vec 0,\vec\alpha_{1})}
\mathcal{K}_{(\vec\alpha_{2},\vec\alpha_{2})}
\times
\mathcal{K}_{(\vec\alpha_{1},\vec\alpha_{1})}
\mathcal{K}_{(2\vec\alpha_{2},\vec\alpha_{2})}
\times
\mathcal{K}_{(2\vec\alpha_{1},\vec\alpha_{1})}
\mathcal{K}_{(3\vec\alpha_{2},\vec\alpha_{2})}
\dots \\ &
\mathcal{K}_{(\vec\alpha_{2},\vec 0)}^{-2}
\mathcal{K}_{(\vec\alpha_{1}+\vec\alpha_{2},\vec 0)}^{-2}
\mathcal{K}_{(\vec\alpha_{1},\vec 0)}^{-2}
\,.
\end{aligned}
\end{equation}
However, this equation is an identity: $\mathcal{K}_{(p\vec\alpha_{1},\vec\alpha_{1})}$ commutes with $\mathcal{K}_{(p\vec\alpha_{2},\vec\alpha_{2})}$, the three purely electric operators commute with each other (two operators commute when symplectic product of their charges is zero); these commuting operators reverse their order precisely at the wall of marginal stability (\ref{weak wall}).
By proving (\ref{WCF3.1}), we have also shown that (\ref{WCF3}) is correct via the substitution of pentagon identity (\ref{pentagon}).

\paragraph{}

Let us now generalise these results to the $SU(n)$ theory.
The approach is very similar.
We can again apply the pentagon identity to the decays of composite dyons (\ref{composite dyon decay}):
\begin{equation}
\label{SU(n) pentagon}
\begin{aligned}
& \mathcal{K}_{
\pm \left(
p\sum_{m=i}^{k}\vec\alpha_{m}+\sum_{l=i+1}^{k}\epsilon_{l}\sum_{m=l}^{k}\vec\alpha_{m}
\,, \
\sum_{m=i}^{k}\vec\alpha_{m}
\right)
}
\\
& \mathcal{K}_{
\pm \left(
\left( p+\sum_{l=i+1}^{k}\epsilon_{l} \right)\sum_{m=k+1}^{j-1}\vec\alpha_{m}+\sum_{l=k+1}^{j-1}\epsilon_{l}\sum_{m=l}^{j-1}\vec\alpha_{m}
\,, \
\sum_{m=k+1}^{j-1}\vec\alpha_{m}
\right)
}
\\
= \,
& \mathcal{K}_{
\pm \left(
\left( p+\sum_{l=i+1}^{k}\epsilon_{l} \right)\sum_{m=k+1}^{j-1}\vec\alpha_{m}+\sum_{l=k+1}^{j-1}\epsilon_{l}\sum_{m=l}^{j-1}\vec\alpha_{m}
\,, \
\sum_{m=k+1}^{j-1}\vec\alpha_{m}
\right)
}
\\
& \mathcal{K}_{
\pm \left(
p\sum_{m=i}^{j-1}\vec\alpha_{m}+\sum_{l=i+1}^{j-1}\epsilon_{l}\sum_{m=l}^{j-1}\vec\alpha_{m}
\,, \
\sum_{m=i}^{j-1}\vec\alpha_{m}
\right)
}
\\
& \mathcal{K}_{
\pm \left(
p\sum_{m=i}^{k}\vec\alpha_{m}+\sum_{l=i+1}^{k}\epsilon_{l}\sum_{m=l}^{k}\vec\alpha_{m}
\,, \
\sum_{m=i}^{k}\vec\alpha_{m}
\right)
}
\,.
\end{aligned}
\end{equation}

\paragraph{}

Suppose that we have a product $S$ of Kontsevich--Soibelman operators for a given vacuum expectation value.
We want to show that all such products are equal.
In order to do this, let us move the VEV continuously into the region where all composite dyons decay.
For each decay process, the product loses one operator according to (\ref{SU(n) pentagon}), but $S$ remains constant.
When VEV is in the region with no composite dyons, this product simplifies to
\begin{equation}
\prod_{p=-\infty}^{+\infty}
\prod_{i=1}^{r}
\mathcal{K}_{(p\vec\alpha_{i},\vec\alpha_{i})}
\times
\prod_{i=1}^{r}
\mathcal{K}_{(\vec\alpha_{i},\vec 0)}^{-2}
\,.
\end{equation}
We have used that for a given $p$ and any $i$, $\mathcal{K}_{(p\vec\alpha_{i},\vec\alpha_{i})}$ commute with each other; purely electric operators commute.
Therefore, every initial product of operators is equal to this expression.
Putting all pieces together, we recover the wall-crossing formula for any weak-coupling region of the moduli space:
\begin{equation}
\begin{aligned}
\mathcal{O}
\left(
\prod_{p=-\infty}^{+\infty}
\prod_{i=1}^{r}
\mathcal{K}_{(p\vec\alpha_{i},\vec\alpha_{i})}
\times
\prod_{i=1}^{r}
\prod_{j=i+1}^{r}
\mathcal{K}_{\left(
p\sum_{m=i}^{j}\vec\alpha_{m}+\sum_{l=i+1}^{j}\epsilon_{l}\sum_{m=l}^{j}\vec\alpha_{m}
\,, \
\sum_{m=i}^{j}\vec\alpha_{m}
\right)}
\times
\prod_{i=1}^{r}
\mathcal{K}_{(\vec\alpha_{i},\vec 0)}^{-2}
\right)
\\
=
\prod_{p=-\infty}^{+\infty}
\prod_{i=1}^{r}
\mathcal{K}_{(p\vec\alpha_{i},\vec\alpha_{i})}
\times
\prod_{i=1}^{r}
\mathcal{K}_{(\vec\alpha_{i},\vec 0)}^{-2}
=
\mathcal{O}
\left(
\prod_{\gamma\in\Gamma(\vec a)}
\mathcal{K}_{\gamma}
\right)
\end{aligned}
\end{equation}
where $\mathcal{O}$ is the clockwise-ordering operator.
This equation relates the spectra far from every wall of marginal stability, in the region with no composite dyons, and located at arbitrary point in the weak-coupling region (where $\Gamma(\vec a)$ is the set of all particles), respectively.

\paragraph{}

The structure of the walls of marginal stability at strong coupling was discussed in \cite{Taylor, Taylor 2}.
The walls existing at weak coupling extend into the strong-coupling region; these walls have already been described above.
As we move the VEV into the strong-coupling region through these walls, all composite dyons disappear from the spectrum.
Further inside this region, there is a group of walls inside which most BPS particles no longer exist, leaving only a finite spectrum.
It is straightforward to generalise the $SU(2)$ formula (\ref{WCF2}) to this case:
\begin{equation}
\mathcal{K}_{(\vec\alpha_{i},-\vec\alpha_{i})}\mathcal{K}_{(\vec 0,\vec\alpha_{i})}
=
\mathcal{K}_{(\vec 0,\vec\alpha_{i})}\mathcal{K}_{(\vec\alpha_{i},\vec\alpha_{i})}\mathcal{K}_{(2\vec\alpha_{i},\vec\alpha_{i})}
\dots \mathcal{K}_{(\vec\alpha_{i},\vec 0)}^{-2} \dots
\mathcal{K}_{(3\vec\alpha_{i},-\vec\alpha_{i})}\mathcal{K}_{(2\vec\alpha_{i},-\vec\alpha_{i})}\mathcal{K}_{(\vec\alpha_{i},-\vec\alpha_{i})}
\,.
\end{equation}
On the left-hand side, we have obtained the finite spectrum inside these strong-coupling walls.


\section{Semiclassical instanton expansion of the compactified theory}

\paragraph{}

Let us consider the $\mathcal{N}=2$ supersymmetric theory compactified on $\mathbb{R}^{3}\times S^{1}$, with radius of $S^{1}$ being $R$.
Employing the Kontsevich--Soibelman wall-crossing formula, it was suggested in \cite{GMN} that the hyper-K\"ahler metric of the moduli space in these compactified theories is determined by a set of integral equations.
Using these equations, we shall perform semiclassical expansion of the metric in the $SU(n)$ theory at weak coupling.
In addition to the single instanton correction to the moduli space
metric, we shall also extract the two-instanton mixing terms and
demonstrate smoothness of the moduli space metric at the walls of
marginal stability. In the next section these results 
will be compared with first principles calculations.

\paragraph{}

We start by recalling some facts about the moduli space metric and introducing our conventions (see \cite{HKLR} for more details).
The three K\"ahler forms $\omega_{1}$, $\omega_{2}$, $\omega_{3}$ corresponding to the hyper-K\"ahler metric can be rewritten as a single form depending on the complex parameter $\zeta$, introduced above:
\begin{equation}
\omega(\zeta) = -\frac{i}{2\zeta}\omega_{+}+\omega_{3}-\frac{i\zeta}{2}\omega_{-}
\end{equation}
where we have introduced $\omega_{\pm}=\omega_{1}\pm i\omega_{2}$.
$\omega_{3}$ is related to the metric of the moduli space via
\begin{equation}
\label{K form}
\omega_{3} = i \, \frac{\partial^{2}K}{\partial z^{a}\partial z^{\bar b}} \, dz^{a}\wedge dz^{\bar b}
\,,
\quad
g = 2 \, \frac{\partial^{2}K}{\partial z^{a}\partial z^{\bar b}} \, dz^{a}dz^{\bar b} = 2 \, g_{a\bar b} \, dz^{a}dz^{\bar b}
\,,
\end{equation}
where $K$ is the corresponding K\"ahler potential.
For a gauge theory of rank $r$ (in our case, $r=n-1$), the symplectic form can be expressed in terms of Darboux coordinates 
\cite{GMN}:
\begin{equation}
\label{symplectic form}
\omega(\zeta) = -\frac{1}{4\pi^{2}R} \, \sum_{I=1}^{r} \,
\frac{d\mathcal{X}_{e}^{I}(\zeta)} {\mathcal{X}_{e}^{I}(\zeta)}
\wedge
\frac{d\mathcal{X}_{m\,I}(\zeta)}{\mathcal{X}_{m\,I}(\zeta)}
\end{equation}
The semiflat metric (i.e., the metric before including corrections coming from the BPS particles wrapping around the compactified dimension) can be obtained by defining Darboux coordinates as
\begin{equation}
\label{coordinate semiflat}
\mathcal{X}_{\gamma}^{\rm sf}(\zeta) = \exp\left( \pi R \zeta^{-1}Z_{\gamma}+i\theta_{\gamma}+\pi R\zeta\bar{Z}_{\gamma} \right)
\end{equation}
where $\theta_{\gamma}=\vec\theta_{e}\vec\gamma_{e}+\vec\theta_{m}\vec\gamma_{m}$, and $\vec\theta_{e}$ and $\vec\theta_{m}$ are the Wilson loops and dual photons appearing after compactifying the theory along $S^{1}$.
The semiflat metric is the approximate form of the metric at very large $R$ where instanton contributions from BPS states of the four-dimensional theory can be neglected.
In the following, we will need the general formula for the Darboux coordinates which holds for all values of $R$.

\paragraph{}

Let us briefly review the method of finding $\mathcal{X}_{\gamma}(\zeta)$ \cite{GMN} by using Kontsevich--Soibelman operators defined in the previous section.
The method states that $\mathcal{X}_{\gamma}(\zeta)$ (for any $\gamma$) are discontinuous along every ray $l$ which is aligned with one or more BPS rays $l_{\gamma'}$ (\ref{BPS ray}).
Explicitly, the jump is given as
\begin{equation}
\label{discontinuity}
\mathcal{X}_{\gamma}^{\text{cw}(l)}(\zeta) = S_{l} \, \mathcal{X}_{\gamma}^{\text{ccw}(l)}(\zeta)
\,,
\quad
S_{l} = \prod_{\gamma'\in\Gamma(\vec a): \, l_{\gamma'}=l} \mathcal{K}_{\gamma'}^{\Omega(\gamma',\vec a)}
\end{equation}
where $\mathcal{X}_{\gamma}^{\text{cw}(l)}(\zeta) $ and $\mathcal{X}_{\gamma}^{\text{ccw}(l)}(\zeta)$ denote $\mathcal{X}_{\gamma}(\zeta)$ as it approaches $l$ clockwise and counterclockwise, respectively.
When $\vec a$ does not belong to a wall of marginal stability, the discontinuities above simplify to
\begin{equation}
\mathcal{X}_{\gamma}^{\text{cw}(l_{\gamma'})}(\zeta) = \mathcal{K}_{\gamma'}^{\Omega(\gamma',\vec a)} \, \mathcal{X}_{\gamma}^{\text{ccw}(l_{\gamma'})}(\zeta)
\,,
\quad
\gamma'\in\Gamma(\vec a)
\,.
\end{equation}
Defining the jumps and knowing the semiflat behaviour of the Darboux coordinates turns out to be enough to recover their values for any $\zeta$.
In general, the set of integral equations for $\mathcal{X}_{\gamma}(\zeta)$ has the form \cite{GMN}
\begin{equation}
\label{RH general}
\mathcal{X}_{\gamma}(\zeta) =
\mathcal{X}_{\gamma}^{\rm sf}(\zeta)
\exp\left(
\frac{1}{4\pi i} \sum_{l}
\int_{l} \frac{d\zeta'}{\zeta'} \frac{\zeta'+\zeta}{\zeta'-\zeta}
\log\frac{\mathcal{X}_{\gamma}(\zeta')}{(S_{l} \, \mathcal{X}_{\gamma})(\zeta')}
\right)
\,.
\end{equation}
The wall-crossing formulae constructed in the previous section show that all $S_{l}$ in this expression are invariant when $\vec a$ crosses walls of marginal stability.
This ensures smoothness of the metric.
The solution of the problem can be re-expressed as a set of $2(n-1)$ integral equations using our conventions for Kontsevich--Soibelman operators:
\begin{equation}
\label{RH}
\mathcal{X}_{\gamma}(\zeta) =
\mathcal{X}_{\gamma}^{\rm sf}(\zeta)
\exp\left(
-\frac{1}{2\pi i} \sum_{\gamma'\in\Gamma(\vec a)}
\Omega(\gamma',\vec a) \langle\gamma,\gamma'\rangle
\int_{l_{\gamma'}} \frac{d\zeta'}{\zeta'} \frac{\zeta'+\zeta}{\zeta'-\zeta}
\log\left( 1-\sigma(\gamma')\mathcal{X}_{\gamma'}(\zeta') \right)
\right)
\end{equation}
where symplectic product and quadratic refinement are defined in (\ref{symplectic product}) and (\ref{quadratic refinement}).
Equation (\ref{RH}), in principle, allows one to construct the metric of the moduli space if the spectrum is known.

\paragraph{}

At this point, we take the weak-coupling limit and approximate the solution of (\ref{RH}) by performing iterations using $\mathcal{X}_{\gamma'}^{\rm sf}(\zeta')$ as the initial approximation for $\mathcal{X}_{\gamma'}(\zeta')$.
The corrected symplectic form will be approximated as
\begin{equation}
\omega(\zeta) \approx \omega^{\rm sf}(\zeta)+\omega^{\rm P}(\zeta)+\omega^{\rm NP}(\zeta)
\end{equation}
where $\omega^{\rm P}(\zeta)$ denotes perturbative corrections from $W$ bosons, $\omega^{\rm NP}(\zeta)$ denotes non-perturbative corrections from monopoles and dyons.
Analogously to the approximation used in \cite{CDP}, our first step is to find the perturbative contributions to the Darboux coordinates: we decompose the perturbatively corrected coordinates as $\mathcal{X}_{\gamma}^{(0)}(\zeta)=\mathcal{X}_{\gamma}^{\rm sf}(\zeta)\mathcal{D}_{\gamma}(\zeta)$ where $\mathcal{D}_{\gamma}(\zeta)$ will be related to the one-loop determinants in the semicalssical calculation.
At leading order, the electric components remain unchanged:
\begin{equation}
\mathcal{X}_{(\vec\gamma_{e},\vec 0)}^{(0)}(\zeta) =
\mathcal{X}_{(\vec\gamma_{e},\vec 0)}^{\rm sf}(\zeta) \mathcal{D}_{(\vec\gamma_{e},\vec 0)}(\zeta) =
\mathcal{X}_{(\vec\gamma_{e},\vec 0)}^{\rm sf}(\zeta)
\,,
\quad
\forall \, \vec\gamma_{e}
\,,
\end{equation}
whereas the magnetic components receive corrections:
\begin{equation}
\mathcal{X}_{(\vec 0,\vec\gamma_{m})}^{(0)}(\zeta) =
\mathcal{X}_{(\vec 0,\vec\gamma_{m})}^{\rm sf}(\zeta) \mathcal{D}_{(\vec 0,\vec\gamma_{m})}(\zeta)
\,,
\quad
\forall \, \vec\gamma_{m}
\,,
\end{equation}
\begin{equation}
\label{D factor}
\begin{aligned}
\log\mathcal{D}_{(\vec 0,\vec\gamma_{m})}(\zeta) =
\frac{1}{\pi i} \sum_{\vec\alpha_{A}\in\Phi_{+}} \vec\gamma_{m}\vec\alpha_{A}
\left(  
\int_{l_{(\vec\alpha_{A},\vec 0)}} \frac{d\zeta'}{\zeta'} \frac{\zeta'+\zeta}{\zeta'-\zeta}
\log\left( 1-\mathcal{X}_{(\vec\alpha_{A},\vec 0)}^{\rm sf}(\zeta') \right)
\right.
\\
\left.
-\int_{l_{(-\vec\alpha_{A},\vec 0)}} \frac{d\zeta'}{\zeta'} \frac{\zeta'+\zeta}{\zeta'-\zeta}
\log\left( 1-1/\mathcal{X}_{(\vec\alpha_{A},\vec 0)}^{\rm sf}(\zeta') \right)
\right)
\end{aligned}
\end{equation}
where $l_{(\pm\vec\alpha_{A},\vec 0)}$ means integrating from zero to infinity along the BPS ray $\vec\alpha_{A}\vec a/\zeta'\in\mathbb{R}_{\mp}$ in the $\zeta'$ plane.
$\mathcal{X}^{\rm sf}$ for electric charges is given by
\begin{equation}
\mathcal{X}_{(\vec\alpha,\vec 0)}^{\rm sf}(\zeta) = \exp\left( \pi R\frac{\vec\alpha\vec a}{\zeta}+i\vec\alpha\vec\theta_{e}+\pi R\vec\alpha\bar{\vec a}\zeta \right)
\,.
\end{equation}

Rotating the contours of integration via introducing $y=-\zeta'/\exp(i\phi_{W_{A}})$ where $\phi_{W_{A}}=\arg(\vec\alpha_{A}\vec a)$ in the first term and $1/y=-\zeta'/\exp(-i\phi_{W_{A}})$ in the second term, we rewrite (\ref{D factor}) as
\begin{equation}
\label{D factor 2}
\begin{aligned}
\log\mathcal{D}_{(\vec 0,\vec\gamma_{m})}(\zeta) =
\frac{1}{\pi i} \sum_{\vec\alpha_{A}\in\Phi_{+}} \vec\gamma_{m}\vec\alpha_{A}
\int_{0}^{+\infty} \frac{dy}{y}
\left( 
\frac{y-\zeta \, e^{-i\phi_{W_{A}}}}{y+\zeta \, e^{-i\phi_{W_{A}}}}
\log\left( 1-e^{-\pi R |\vec\alpha_{A}\vec a| (y+1/y)+i\vec\alpha_{A}\vec\theta_{e}} \right)
\right.
\\
\left.
-\frac{y-\zeta^{-1}e^{i\phi_{W_{A}}}}{y+\zeta^{-1}e^{i\phi_{W_{A}}}}
\log\left( 1-e^{-\pi R |\vec\alpha_{A}\vec a| (y+1/y)-i\vec\alpha_{A}\vec\theta_{e}} \right)
\right)
\,.
\end{aligned}
\end{equation}
Note that this expression is real if and only if $|\zeta|=1$.
The perturbative corrections are given as
\begin{equation}
\omega^{\rm P}(\zeta) = -\frac{1}{4\pi^{2}R} \sum_{I=1}^{r}
d\log\mathcal{X}_{e}^{I\,\rm sf}(\zeta)
\wedge
d\log\mathcal{D}_{\vec E_{I}}(\zeta)
\end{equation}
where $I$-th component of $\vec E_{I}$ is 1, and all other components are 0.


\subsection{Non-perturbative corrections and smoothness of the metric}

\paragraph{}

The non-perturbative corrections to the metric are given by the
following iterative expansion of the integral equation:
\begin{equation}
\begin{aligned}
\label{RH correction}
\delta\log\mathcal{X}_{\gamma}^{(n)}(\zeta) & =
-\frac{1}{2\pi i} \sum_{\gamma'\in\tilde\Gamma(\vec a)}
\Omega(\gamma',\vec a) \langle\gamma,\gamma' \rangle
\int_{l_{\gamma'}} \frac{d\zeta'}{\zeta'} \frac{\zeta'+\zeta}{\zeta'-\zeta}
\log\left( 1-\sigma(\gamma) \, \mathcal{X}_{\gamma'}^{(n-1)}(\zeta') \right)
\,,
\\
\mathcal{X}^{(n)}_{\gamma}(\zeta) & = \mathcal{X}_{\gamma'}^{(0)}
\exp\left( \delta\log\mathcal{X}_{\gamma'}^{(n)}(\zeta) \right)
\,,
\quad
n \in \mathbb{N}
\,, 
\end{aligned} 
\end{equation}
where $\tilde\Gamma(\vec a)$ includes only the non-perturbative BPS spectrum, i.e., monopoles, dyons, and their anti-particles. The superscript $n$ in $\mathcal{X}^{(n)}_{\gamma}(\zeta)$ should be understood as keeping up to $n$-instanton terms in the series expansion.
This process is iterative and will be illustrated for $n=2$: we will find the contributions for one and two dyons.
Of course, the smoothness property of $\mathcal{X}_{\gamma}(\zeta)$ is built in by construction \cite{GMN}, here our manipulation in instanton expansion merely makes this explicit and suitable for the semiclassical instanton checks.

\paragraph{}

Using (\ref{RH correction}), we can write out the explicit expression for $\mathcal{X}_{\gamma'}^{(1)}(\zeta')$:
\begin{equation}
\mathcal{X}_{\gamma'}^{(1)}(\zeta') = \mathcal{X}_{\gamma'}^{(0)}(\zeta')
\exp\left(
\frac{1}{2\pi i} \sum_{\gamma''\in\tilde\Gamma(\vec a)}
\Omega(\gamma'',\vec a) \langle\gamma',\gamma''\rangle
\int_{l_{\gamma''}} \frac{d\zeta''}{\zeta''} \frac{\zeta''+\zeta'}{\zeta''-\zeta'}
\sum_{l=1}^{+\infty} \frac{1}{l} \left( \sigma(\gamma'')\mathcal{X}_{\gamma''}^{(0)}(\zeta'') \right)^{l} \right)
\,,
\end{equation}
in our two-instanton calculation, it is sufficient to set $l=1$.
Furthermore, since $|\delta\log\mathcal{X}^{(1)}_{\gamma'}(\zeta')|$ is small, we make the following approximation:
\begin{equation}
\label{Darboux 1}
\mathcal{X}^{(1)}_{\gamma'}(\zeta') \approx
\mathcal{X}^{(0)}_{\gamma'}(\zeta')
\left( 1+\delta\log\mathcal{X}^{(1)}_{\gamma'}(\zeta') \right)
\,,
\end{equation}
where we are only keeping up to two instanton terms in the expansion.
Higher order terms in $\exp(\delta\log\mathcal{X}^{(1)}(\zeta'))$ expansion will contribute to $n>2$ instanton terms.
The explicit non-perturbative corrections to the symplectic form up to this order are
\begin{equation}
\label{2 dyons correction}
\omega^{\rm NP}(\zeta) \approx \omega^{\rm NP(1)}(\zeta)+\omega^{\rm NP(2)}(\zeta)+\omega^{\rm NP(\tilde 2)}(\zeta)
\end{equation}
where
\begin{equation}
\begin{aligned}
\omega^{\rm NP(1)}(\zeta) + \omega^{\rm NP(2)}(\zeta) = -\frac{1}{4\pi^{2}R} \sum_{I=1}^{r}
\left(
d\log\mathcal{X}_{e}^{I\,(0)}(\zeta)
\wedge
d\delta\log\mathcal{X}_{m\,I}^{(2)}(\zeta)
\right.
\\
\left.
+
d\delta\log\mathcal{X}_{e}^{I\,(2)}(\zeta)
\wedge
d\log\mathcal{X}_{m\,I}^{(0)}(\zeta)
\right)
\,,\label{w1w2}
\end{aligned}
\end{equation}
\begin{equation}
\omega^{\rm NP(\tilde 2)}(\zeta) = -\frac{1}{4\pi^{2}R} \sum_{I=1}^{r}
d\delta\log\mathcal{X}_{e}^{I\,(1)}(\zeta)
\wedge
d\delta\log\mathcal{X}_{m\,I}^{(1)}(\zeta)
\,.
\end{equation}
We shall define the explicit expressions for $\omega^{\rm NP (1)}(\zeta)$ and $\omega^{\rm NP(2)}(\zeta)$ momentarily.
We should note here that all terms in $\omega^{\rm NP(\tilde 2)}(\zeta)$ are continuous as they consist of mixing terms between simple dyons which exist everywhere in the weakly coupled region of moduli space.
They do not affect the smoothness property of the metric, and we shall ignore them in this section.

\paragraph{}

In (\ref{w1w2}), $\omega^{\rm NP(1)}(\zeta)$ is a series of instanton terms proportional to $\exp\left( -2\pi kR |Z_{\gamma'}|+ik\theta_{\gamma'} \right)$:
\begin{equation}
\label{symplectic form NP1}
\begin{aligned}
\omega^{\rm NP(1)}(\zeta)
=
-\frac{1}{4\pi^{2}R} \, \frac{1}{2\pi i}
\sum_{\gamma'\in\tilde\Gamma(\vec a)}
\Omega(\gamma',\vec a) \,
\frac{d\mathcal{X}_{\gamma'}^{(0)}(\zeta)}{\mathcal{X}_{\gamma'}^{(0)}(\zeta)} \,
\wedge \,
\int_{l_{\gamma'}} \frac{d\zeta'}{\zeta'} \frac{\zeta'+\zeta}{\zeta'-\zeta}
\\
\sum_{k=1}^{+\infty} \left( \sigma(\gamma') \, \mathcal{X}_{\gamma'}^{(0)}(\zeta') \right)^{k}
d\log\mathcal{X}_{\gamma'}^{(0)}(\zeta')
\,.
\end{aligned}
\end{equation}
Note that $\tilde{\Gamma}(\vec a)$ here consists of both simple and composite dyons, and $k$ is the winding number of the dyon world line over the compactified $S^1$.
For checking that the smoothness of the moduli space metric, it is sufficient to consider only the singly wound dyons $k=1$.
The two-instanton correction $\omega^{\rm NP(2)}(\zeta)$ in (\ref{w1w2}) corresponds to setting $k=1$ and integrating along two BPS rays:
\begin{equation}
\label{symplectic form NP2}
\begin{aligned}
\omega^{\rm NP(2)}(\zeta)
= &
-\frac{1}{4\pi^{2}R} \left( \frac{1}{2\pi i} \right)^{2}
\sum_{\{\gamma',\gamma''\}\subset\tilde\Gamma(\vec a)}
\Omega(\gamma',\vec a) \, \Omega(\gamma'',\vec a) \, \sigma(\gamma') \, \sigma(\gamma'') \, \langle\gamma',\gamma''\rangle
\\
& \frac{d\mathcal{X}_{\gamma'}^{(0)}(\zeta)}{\mathcal{X}_{\gamma'}^{(0)}(\zeta)} \,
\wedge
\int_{l_{\gamma'}}\frac{d\zeta'}{\zeta'} \frac{\zeta'+\zeta}{\zeta'-\zeta} \,
\int_{l_{\gamma''}} \frac{d\zeta''}{\zeta''} \frac{\zeta''+\zeta'}{\zeta''-\zeta'} \,
\left(
\mathcal{X}_{\gamma''}^{(0)}(\zeta'') \, d\mathcal{X}_{\gamma'}^{(0)}(\zeta') +
\mathcal{X}_{\gamma'}^{(0)}(\zeta') \, d\mathcal{X}_{\gamma''}^{(0)}(\zeta'')
\right)
\,.
\end{aligned}
\end{equation}
We can decompose (\ref{symplectic form NP1}) and (\ref{symplectic form NP2}) as
\begin{equation}
\omega^{\rm NP(1)}(\zeta) = \sum_{\gamma'\in\tilde\Gamma(\vec a)} \omega_{\gamma'}^{\rm NP(1)}(\zeta)
\,,
\quad
\omega^{\rm NP(2)}(\zeta) = \sum_{\{\gamma',\gamma''\}\subset\tilde\Gamma(\vec a)} \omega_{\gamma',\gamma''}^{\rm NP(2)}(\zeta)
\,.
\end{equation}
Suppose that the VEV crosses the wall where a dyon with charge $\gamma_{1}+\gamma_{2}$ changes its multiplicity by $\Delta\Omega(\gamma_{1}+\gamma_{2},\vec a)$.
To ensure smoothness of the metric, one needs to make sure that
\begin{equation}
\label{continuity condition}
\left(
\lim_{\arg\frac{Z_{\gamma_{2}}}{Z_{\gamma_{1}}}\to 0+} -
\lim_{\arg\frac{Z_{\gamma_{2}}}{Z_{\gamma_{1}}}\to 0-}
\right)
\left(
\omega_{\gamma_{1}+\gamma_{2}}^{\rm NP(1)}(\zeta)+
\omega_{\gamma_{1},\gamma_{2}}^{\rm NP(2)}(\zeta)+
\omega_{\gamma_{2},\gamma_{1}}^{\rm NP(2)}(\zeta)
\right)
= 0
\,.
\end{equation}
This condition imposes a constraint on multiplicities on both sides of the wall.
After finding this constraint, we will see that it is indeed satisfied by our pentagon identities.

\paragraph{}

Let us see how $\omega_{\gamma_{1},\gamma_{2}}^{\rm NP(2)}(\zeta)$  and $\omega_{\gamma_{2},\gamma_{1}}^{\rm NP(2)}(\zeta)$ change when we cross the wall of marginal stability.
We need to identify $\gamma'=\gamma_{1}$, $\gamma''=\gamma_{2}$ or $\gamma'=\gamma_{2}$, $\gamma''=\gamma_{1}$ in (\ref{symplectic form NP2}).
Using the fact that at the wall, $l_{\gamma_{1}}$, $l_{\gamma_{2}}$, and $l_{\gamma_{1}+\gamma_{2}}$ coincide, we see that the jump of $\omega^{\rm NP(2)}(\zeta)$ corresponds to the residue of the second (internal) integral at $\zeta''=\zeta'$ in (\ref{symplectic form NP2}):
\begin{equation}
\begin{aligned}
\left( \lim_{\arg\frac{Z_{\gamma_{2}}}{Z_{\gamma_{1}}}\to 0-} - \lim_{\arg\frac{Z_{\gamma_{2}}}{Z_{\gamma_{1}}}\to 0+} \right)
\left(
\omega_{\gamma_{1},\gamma_{2}}^{\rm NP(2)}(\zeta) +
\omega_{\gamma_{2},\gamma_{1}}^{\rm NP(2)}(\zeta)
\right)
=
- \, \Omega(\gamma_{1},\vec a) \, \Omega(\gamma_{2},\vec a) \, \sigma(\gamma_{1}) \, \sigma(\gamma_{2}) \, 2 \, \langle\gamma_{1},\gamma_{2}\rangle
\\
\frac{1}{4\pi^{2}R} \, \frac{1}{2\pi i} \,
\frac{d\mathcal{X}_{\gamma_{1}+\gamma_{2}}^{(0)}(\zeta)}{\mathcal{X}_{\gamma_{1}+\gamma_{2}}^{(0)}(\zeta)} \,
\wedge
\int_{l_{\gamma_{1}+\gamma_{2}}}\frac{d\zeta'}{\zeta'}\frac{\zeta'+\zeta}{\zeta'-\zeta} \,
d\mathcal{X}_{\gamma_{1}+\gamma_{2}}^{(0)}(\zeta')
\,,
\end{aligned}
\end{equation}
where the jumps of $\omega_{\gamma_{1},\gamma_{2}}^{\rm NP(2)}$ and $\omega_{\gamma_{2},\gamma_{1}}^{\rm NP(2)}$ are equal.
The increment of $\omega^{\rm NP(1)}(\zeta)$ across the wall can be easily seen from (\ref{symplectic form NP1}) setting $\gamma'=\gamma_{1}+\gamma_{2}$:
\begin{equation}
\begin{aligned}
\left( \lim_{\arg\frac{Z_{\gamma_{2}}}{Z_{\gamma_{1}}}\to 0-} - \lim_{\arg\frac{Z_{\gamma_{2}}}{Z_{\gamma_{1}}}\to 0+} \right)
\omega_{\gamma_{1}+\gamma_{2}}^{\rm NP(1)}(\zeta)
=
- \, \Delta\Omega(\gamma_{1}+\gamma_{2},\vec a) \, \sigma(\gamma_{1}+\gamma_{2})
\\
\frac{1}{4\pi^{2}R} \, \frac{1}{2\pi i} \,
\frac{d\mathcal{X}_{\gamma_{1}+\gamma_{2}}^{(0)}(\zeta)}{\mathcal{X}_{\gamma_{1}+\gamma_{2}}^{(0)}(\zeta)} \,
\wedge
\int_{l_{\gamma_{1}+\gamma_{2}}}\frac{d\zeta'}{\zeta'}\frac{\zeta'+\zeta}{\zeta'-\zeta} \,
d\mathcal{X}_{\gamma_{1}+\gamma_{2}}^{(0)}(\zeta')
\,.
\end{aligned}
\end{equation}
Using the relation between two quadratic refinements, $\sigma(\gamma_{1})\,\sigma(\gamma_{2})=(-1)^{2\langle\gamma_{1},\gamma_{2}\rangle}\sigma(\gamma_{1}+\gamma_{2})$, we can see that the continuity condition (\ref{continuity condition}) is equivalent to
\begin{equation}
\Delta\Omega(\gamma_{1}+\gamma_{2},\vec a) =
2 \, \langle\gamma_{1},\gamma_{2}\rangle \,
(-1)^{2\langle\gamma_{1},\gamma_{2}\rangle-1} \,
\Omega(\gamma_{1},\vec a) \,
\Omega(\gamma_{2},\vec a)
\,.
\end{equation}
It is indeed ensured by (\ref{SU(3) pentagon}) and (\ref{SU(n) pentagon}): $\Delta\Omega(\gamma_{1}+\gamma_{2},\vec a)=\Omega(\gamma_{1},\vec a)=\Omega(\gamma_{2},\vec a)=1$, $\langle\gamma_{1},\gamma_{2}\rangle=1/2$.
This allows us to conclude that the moduli space metric remains continuous to the two-instanton order across the WMS where composite dyons decay.
This analysis can be repeated to ensure the smoothness of higher-instanton mixing terms across WMS by expanding systematically the higher $\mathcal{X}^{(n)}_{\gamma}(\zeta)$ terms.


\subsection{Saddle-point approximation of the metric}

\paragraph{}

Knowing the general expressions for one and two-instanton corrections (\ref{symplectic form NP1}, \ref{symplectic form NP2}), we can now extract the moduli space metric using the saddle-point approximation~\footnote{
Explicitly, the approximation we are using is
\begin{equation}
\label{saddle point}
\int_{a}^{b} e^{f(x)}dx \approx \sqrt\frac{2\pi}{|f''(x_{0})|} \, e^{f(x_{0})}
\end{equation}
for $f(x)$ having sharp peak at $x=x_{0}$, $a<x_{0}<b$.
}.
To approximate (\ref{symplectic form NP2}), this method can only be used far from the walls, where the integrands do not have poles near the contour of integration.
For the terms in (\ref{symplectic form NP1}), the peak is at $\zeta'=-Z_{\gamma'}/|Z_{\gamma'}|=-e^{i\phi_{\gamma'}}$ (where $\phi_{\gamma'}$ is the complex argument of $Z_{\gamma'}$).
Proceding as in \cite{CDP}, we obtain
\begin{equation}
\label{symplectic form NP1 saddle}
\begin{aligned}
\omega^{\rm NP(1)}(\zeta)
= &
\frac{i}{8\pi^{2}}
\sum_{\gamma'\in\tilde\Gamma(\vec a)}
\sum_{k=1}^{+\infty}
\left( \mathcal{D}_{\gamma'}(-e^{i\phi_{\gamma'}}) \right)^{k} \,
\frac{1}{\sqrt{kR|Z_{\gamma'}|}} \exp\left( -2\pi kR |Z_{\gamma'}|+ik\theta_{\gamma'} \right)
\\
& \frac{d\mathcal{X}_{\gamma'}^{\rm sf}(\zeta)}{\mathcal{X}_{\gamma'}^{\rm sf}(\zeta)}
\wedge
\left(
|Z_{\gamma'}| \left( \frac{dZ_{\gamma'}}{Z_{\gamma'}}-\frac{d\bar{Z}_{\gamma'}}{\bar{Z}_{\gamma'}} \right)
-\left( \frac{dZ_{\gamma'}}{\zeta}-\zeta d\bar{Z}_{\gamma'} \right)
\right)
\,,
\end{aligned}
\end{equation}
where the global definition of $\vec\theta_{m}$ leads to the shift $\vec\theta_{m}\to\vec\theta_{m}+\re\hat\tau_{\rm eff} \, \vec\theta_{e}$ in $\theta_{\gamma}$ in order to define it consistently at infinity.
Now, let us approximate the two-instanton terms, (\ref{symplectic form NP2}).
Using the same saddle-point approximation (\ref{saddle point}) in both integrals here (the maxima of the integrands are at $\zeta'=-Z_{\gamma'}/|Z_{\gamma'}|=-e^{i\phi_{\gamma'}}$ and $\zeta''=-Z_{\gamma''}/|Z_{\gamma''}|=-e^{i\phi_{\gamma''}}$), we see that the two-instanton terms are proportional to $\exp\left( -2\pi R(|Z_{\gamma'}|+|Z_{\gamma''}|) \right)$, correctly reproducing the two-instanton action.
At weak coupling, when masses of all dyons are large, (\ref{symplectic form NP2}) gives next order corrections with respect to (\ref{symplectic form NP1 saddle}).
Computing the third component of the symplectic form, $\omega_{3}=\left( (\omega(i)+\omega(-i) \right)/2$, we express the contribution for two dyons in terms of their central charges:
\begin{equation}
\label{symplectic form 3 NP2}
\begin{aligned}
\omega_{3}^{\rm NP(2)}
= &
-\frac{1}{4\pi^{2}R} \left( \frac{1}{2\pi i} \right)^{2}
\sum_{\{\gamma',\gamma''\}\subset\tilde\Gamma(\vec a)}
\mathcal{S}_{\gamma',\gamma''} \,
\frac{1}{R\sqrt{|Z_{\gamma'}Z_{\gamma''}|}} \,
\sigma(\gamma') \, \sigma(\gamma'') \,
\frac{\vec\gamma_{m}' \, \vec\gamma_{e}'' - \vec\gamma_{e}' \, \vec\gamma_{m}''}{2} \,
\frac{e^{i\phi_{\gamma'}}+e^{i\phi_{\gamma''}}}{e^{i\phi_{\gamma'}}-e^{i\phi_{\gamma''}}}
\\
&
\left(
\frac{e^{i\phi_{\gamma'}}-i}{e^{i\phi_{\gamma'}}+i}
\left( i\pi R\left( -dZ_{\gamma'}+d\bar Z_{\gamma'} \right)+id\theta_{\gamma'} \right)
+
\frac{e^{i\phi_{\gamma'}}+i}{e^{i\phi_{\gamma'}}-i}
\left( i\pi R\left( dZ_{\gamma'}-d\bar Z_{\gamma'} \right)+id\theta_{\gamma'} \right)
\right)
\wedge
\\
&
\left(
-\pi R\left( e^{-i\phi_{\gamma'}}dZ_{\gamma'}+e^{i\phi_{\gamma'}}d\bar Z_{\gamma'} \right)
-\pi R\left( e^{-i\phi_{\gamma''}}dZ_{\gamma''}+e^{i\phi_{\gamma''}}d\bar Z_{\gamma''} \right) +
id\theta_{\gamma'+\gamma''}
\right)
\end{aligned}
\end{equation}
where the factor describing dyon actions and non-zero modes determinants is
\begin{equation}
\mathcal{S}_{\gamma',\gamma''} =
\mathcal{D}_{\gamma'}(-e^{i\phi_{\gamma'}}) \, \mathcal{D}_{\gamma''}(-e^{i\phi_{\gamma''}})
\exp\left( -2\pi R(|Z_{\gamma'}|+|Z_{\gamma''}|)+i\theta_{\gamma'+\gamma''} \right)
\,.
\end{equation}
Note that (\ref{symplectic form 3 NP2}) is applicable only far from the walls of marginal stability: it diverges at the wall as the contour of integration passes through a pole, where our saddle point approximation cannot be used. 

\paragraph{}

Let us now extract the dominant metric components, $g_{a^{I}\bar a^{J}}$, from these symplectic forms.
At weak coupling, all central charges can be approximated as
\begin{align}
Z_{\gamma} & = \vec\gamma_{e}\vec a+\vec\gamma_{m}\hat\tau_{\rm eff}\vec a
\,,
\\
\hat\tau_{\rm eff} & \simeq \frac{i}{\pi} \sum_{\vec\alpha_{A}\in\Phi_{+}}
\vec\alpha_{A}\otimes\vec\alpha_{A}
\log\left( \frac{\vec\alpha_{A}\vec a}{\Lambda} \right)^{2}
\,.
\end{align}
Further, everywhere, except the exponents, we can approximate central charges for dyons as $Z_{\gamma}\simeq\vec\gamma_{m}(i\im\hat\tau_{\rm eff})\vec a$, $\bar Z_{\gamma}\simeq -\vec\gamma_{m}(i\im\hat\tau_{\rm eff})\bar{\vec a}$.
For the symplectic product of central charges, we have
\begin{equation}
dZ_{\gamma'}\wedge d\bar Z_{\gamma''} \simeq
(\vec\gamma_{m}'\im\hat\tau_{\rm eff})_{I} \, (\vec\gamma_{m}''\im\hat\tau_{\rm eff})_{J} \,
da^{I}\wedge d\bar a^{J}
\,,
\quad
\im\hat\tau_{\rm eff} \simeq \frac{2}{\pi} \sum_{\vec\alpha_{A}\in\Phi_{+}}
\vec\alpha_{A}\otimes\vec\alpha_{A} \,
\log\left| \frac{\vec\alpha_{A}\vec a}{\Lambda} \right|
\,.
\end{equation}
The resulting correction for single dyons is
\begin{equation}
\label{metric NP1}
\begin{aligned}
g_{a^{I}\bar a^{J}}^{\rm NP(1)}
= &
\frac{1}{4\pi}
\sum_{\gamma'\in\tilde\Gamma(\vec a)}
\sum_{k=1}^{+\infty}
\left( \mathcal{D}_{\gamma'}(-e^{i\phi_{\gamma'}}) \right)^{k}
\exp\left( -2k\pi R |Z_{\gamma'}|+ik\theta_{\gamma'} \right)
\sqrt{\frac{R}{k|Z_{\gamma'}|}}
\\
&
(\vec\gamma_{m}'\im\hat\tau_{\rm eff})_{I} \, (\vec\gamma_{m}'\im\hat\tau_{\rm eff})_{J}
\,.
\end{aligned}
\end{equation}
After some tedious but straightforward calculations, we obtain the dominant components of the moduli space metric coming from pairs of dyons:
\begin{equation}
\label{metric NP2}
\begin{aligned}
g_{a^{I}\bar a^{J}}^{\rm NP(2)}
= &
-\frac{1}{16\pi^{2}}
\sum_{\{\gamma',\gamma''\}\subset\tilde\Gamma(\vec a)}
\mathcal{S}_{\gamma',\gamma''} \,
\frac{1}{\sqrt{|Z_{\gamma'}Z_{\gamma''}|}} \,
\sigma(\gamma') \, \sigma(\gamma'') \,
i(\vec\gamma_{m}' \, \vec\gamma_{e}'' - \vec\gamma_{e}' \, \vec\gamma_{m}'') \,
\frac{e^{i\phi_{\gamma'}}+e^{i\phi_{\gamma''}}}{e^{i\phi_{\gamma'}}-e^{i\phi_{\gamma''}}}
\\
&
\left(
2 \, (\vec\gamma_{m}'\im\hat\tau)_{I} \, (\vec\gamma_{m}'\im\hat\tau)_{J} +
\frac{\exp(i\phi'')}{\cos\phi'} \, (\vec\gamma_{m}'\im\hat\tau)_{I} \, (\vec\gamma_{m}''\im\hat\tau)_{J} +
\right.
\\
& \qquad
\left.
\frac{\exp(-i\phi'')}{\cos\phi'} \, (\vec\gamma_{m}''\im\hat\tau)_{I} \, (\vec\gamma_{m}'\im\hat\tau)_{J}
\right)
\,.
\end{aligned}
\end{equation}
The reality condition for these expressions can be checked using the fact that these summations are symmetric under $\gamma'\to-\gamma'$, $\gamma''\to-\gamma''$.

\paragraph{}

Let us express the perturbative one-loop factor extracted from \cite{GMN}, i.e., $\mathcal{D}_{\gamma}(\zeta)$ in (\ref{D factor 2}), explicitly; then, we will explain how it can be reproduced from semiclassical analysis.
First, we notice that in the semiclassical limit, the phase $\phi_{\gamma}$ is given via
\begin{equation}
\exp(i\phi_{\gamma}) =
\frac{(\gamma_{e\,I}+\tau_{{\rm eff}\,IJ}\gamma_{m}^{J})a^{I}}{|(\gamma_{e\,I}+\tau_{{\rm eff}\,IJ}\gamma_{m}^{J})a^{I}|} \simeq
\frac{\tau_{{\rm eff}\,IJ}\gamma_{m}^{J}a^{I}}{|\tau_{{\rm eff}\,IJ}\gamma_{m}^{J}a^{I}|}
\,.
\end{equation}
This is a non-trivial generalisation of the 
rank one case where $\exp(i\phi_{\gamma})\simeq ia/|a|$: in the $SU(n)$ case, even in the semiclassical limit, the phase $\phi_{\gamma}$ remains different for monopoles and dyons charged under different roots, and so, we need to carefully re-evaluate the one-loop factors.

\paragraph{}

For a given monopole $\gamma_{A}=(\vec 0,\vec\alpha_{A})$ charged under root $\vec\alpha_{A}$ (simple or composite), we can split the summation over different $W$ bosons into the term where the boson is charged under $\vec\alpha_{A}$ and all other terms where the boson is charged under $\vec\alpha_{B\ne A}$ roots.
We can then rewrite $\log\mathcal{D}_{(\vec\gamma_{e},\vec\gamma_{m})}(\zeta)$ (for any charges $(\vec\gamma_{e},\vec\gamma_{m})$) at the saddle point $\zeta=-e^{i\phi_{\gamma_{A}}}$ as
\begin{equation}
\label{D factor 3}
\log\mathcal{D}_{(\vec\gamma_{e},\vec\gamma_{m})} (-e^{i\phi_{\gamma_{A}}}) =
\log\mathcal{D}_{(\vec\gamma_{e},\vec\gamma_{m}),A} (-e^{i\phi_{\gamma_{A}}})+
\sum_{B\ne A} \log\mathcal{D}_{(\vec\gamma_{e},\vec\gamma_{m}),B} (-e^{i\phi_{\gamma_{A}}})
\,.
\end{equation}
Introducing $y=e^{t}$ in (\ref{D factor 2}), we re-express the first term (coming from the $W_{A}$ boson and its antiparticle, which are also charged under $\vec\alpha_{A}$):
\begin{equation}
\label{D factor simple}
\begin{aligned}
\log\mathcal{D}_{(\vec\gamma_{e},\vec\gamma_{m}),A} (-e^{i\phi_{\gamma_{A}}}) & =
\frac {2 \, \vec\alpha_{A}\vec\gamma_{m}}{\pi} \int_{0}^{+\infty} \frac{dt}{\cosh t}
\\
& \left( \log\left( 1-e^{-2\pi R |Z_{W_{A}}|\cosh t+i\theta_{W_{A}}} \right) + \log\left( 1-e^{-2\pi R |Z_{W_{A}}|\cosh t-i\theta_{W_{A}}} \right) \right)
\,.
\end{aligned}
\end{equation}
This term is analogous to the $SU(2)$ one-loop factor evaluated in \cite{CDP}. 
For generic gauge group, it also has contributions from other roots $\vec\alpha_{B\ne A}$.
To calculate them, we set $\vec\alpha_{B}$ in the summation in (\ref{D factor 2}) and substitute $y=e^{t}$ as above, then, at the saddle point $\zeta=-e^{i\phi_{\gamma_{A}}}$, we express these terms in terms of complex phases $\phi_{\gamma_{A}}$ and $\phi_{W_{B}}$:
\begin{equation}
\label{D factor mixed}
\begin{aligned}
\log\mathcal{D}_{(\vec\gamma_{e},\vec\gamma_{m}),B} (-e^{i\phi_{\gamma_{A}}}) & =
\frac{2 \, \vec\alpha_{A}\vec\gamma_{m}}{\pi} \int_{0}^{+\infty} dt \, \rho(t,\Delta\phi)
\\
& \left( \log\left( 1-e^{-2\pi R |Z_{W_{B}}|\cosh t+i\theta_{W_{B}}} \right)+\log\left( 1-e^{-2\pi R |Z_{W_{B}}|\cosh t-i\theta_{W_{B}}} \right) \right)
\end{aligned}
\end{equation}
where the integration kernel is given by
\begin{equation}
\label{D factor kernel}
\rho(t,\Delta\phi) =
\frac{\cosh t\cos\Delta\phi}{\cosh^{2}t-\sin^{2}\Delta \phi} =
\frac{\cos\Delta\phi}{2}\left( \frac{1}{\cosh t-\sin\Delta\phi}+\frac{1}{\cosh t+\sin\Delta\phi} \right)
\end{equation}
with $\Delta\phi=\phi_{W_{B}}-\phi_{\gamma_{A}}+\frac{\pi}{2}=\phi_{W_{B}}-\phi_{W_{A}}$ (the case we are dealing with is $(\vec\gamma_{e},\vec\gamma_{m})=\gamma_{A}$).
In three dimensions, $(\re\vec a,\im\vec a,\vec\theta_{e}/2\pi R)$ form a vector of enhanced $SO(3)$ triplets, and the one-loop factor should be invariant under such rotations.
In the next section, we will use this property to match this expression with the semiclassical result for non-zero mode fluctuations.


\section{Instanton calculus in compactified gauge theories}

\paragraph{}

In this section, we shall discuss the semiclassical field theory computation of the moduli space metric and four-fermion correlation function in the instanton background, 
which serve as non-trivial checks for the instanton corrections to the moduli space metric obtained in the previous section.
Some of the details present here were given in \cite{CDP}, to keep the discussion concise, we shall refer readers to that reference wherever appropriate.


\subsection{Structure of correlation function in instanton background}
\label{sec: instanton correlator}

\paragraph{}

In our compactified $\mathcal{N}=2$ $SU(n)$ gauge theory on ${\mathbb R}^{3}\times S^{1}$, the four-fermion correlation function in the $P$-monopole background is given as~\footnote{
The topological charge $P=\sum_{B=1}^{N-1} m_{B}$, where $m_{B}$ is the magnetic charge under a simple root $\alpha_{B}$, so that for a monopole charged under composite root $\alpha_{1}+\alpha_{2}$, we have $P=2$ and so on.
}:
\begin{equation}
\begin{aligned}
\mathcal{G}_{4}^{(P)}(y_{1},y_{2},y_{3},y_4) = \int [d\mu_{B}^{(P)}][d\mu_F^{(P)}] \, {\mathcal{R}}^{(P)}
\prod^{2}_{A=1}\rho_{1}^{(P)}(y_{2A-1})\rho_{2}^{(P)}(y_{2A})
\\
\exp\left(-\int^{2\pi R}_{0}dx_{4}L_{QM}^{(P)} - S_{\rm Mon.}^{(P)}\right)
\label{Def4Fermi}
\end{aligned}
\end{equation}
where $[d\mu_{B}^{(P)}]$ and $[d\mu_F^{(P)}]$ are the bosonic and fermionic integration measures for the zero mode flucutations in the $P$-monopole background, which involves integrating over the $P$-monopole moduli space in $SU(n)$ gauge theory.
The additional factor $\mathcal{R}^{(P)}$ is the one-loop determinant which sums over all other non-zero mode fluctuations in the monopole background,
which will be discussed extensively in the next section.
The moduli space metric for general partition of magnetic charges $\{m_{B}\}\,,~ P=\sum_{B=1}^{N-1} m_{B}$ is unknown, however, for a special configuration where $m_{B}=1\,, \forall B$, i.e., for simple monopoles, the exact moduli space metric was successfully obtained in \cite{LWY}, and allow for semiclassical quantizations.
We will mainly focus on such configuration.
It is important to note that the non-trivial interaction terms in the monopole moduli space metric are proportional to the Cartan matrix, therefore, in such specific configuration, only pair-wise electromagnetic interactions between simple monopoles charged under adjacent simple roots in Dynkin diagram give non-trivial contributions.

\paragraph{}

At weak coupling, $P$-monopoles become very massive, their low-energy semiclassical dynamics is governed by the supersymmetric quantum mechanics over the $P$-monopole moduli space. We can separate the supersymmetric lagrangian $L_{\rm QM}^{(P)}$ into two parts: one part corresponds to the motion of the centre of mass, the other part corresponds to the motion of $P$-monopoles in the relative moduli space where the interactions takes place, that is:
\begin{equation}
L_{\rm QM}^{(P)}=L_{\rm COM}+L_{\rm Rel.}^{(P)}\,.\label{SLQM}
\end{equation}
We can similarly separate the four-fermion correlation $\mathcal{G}_{4}^{(P)}$ into two parts, $\mathcal{G}_{4}^{(P)}=\mathcal{G}_{\rm COM}^{(4)}\times {\mathcal Z}^{(P)}$, where
\begin{equation}
\begin{aligned}
\mathcal{G}_{\rm COM}^{(4)}(y_{1},y_{2},y_{3},y_4)
=\int [d^{3}X(x^{4})][d\Phi(x^{4})][d^{4}\Psi(x^{4})]{\mathcal{R}}^{(P)}
\prod^{2}_{A=1}\rho_{1}^{(P)}(y_{2A-1})\rho_{2}^{(P)}(y_{2A})
\\
\exp\left(-\int^{2\pi R}_{0}dx_{4}L_{\rm COM} - S_{\rm Mon.}^{(P)}\right)\,,
\label{COM4Fermi}
\end{aligned}
\end{equation}
\begin{equation}
{\mathcal Z}^{(P)}=\int [d^{4P-4}\mu_{B}^{\rm Rel.}][d^{4P-4}\mu_{F}^{\rm Rel.}]\exp\left(-\int^{2\pi R}_{0}dx^{4} L_{\rm Rel.}^{(P)}\right)\,.
\end{equation}
In $\mathcal{G}_{\rm COM}^{(4)}(y_{1},y_{2},y_{3},y_4)$, the three bosonic zero modes $X_{1,2,3}$ correspond to the centre of mass coordinates for the $P$-monopole configuration, and $\Phi$ corresponds to the angle of overall global $U(1)$ rotation. They are accompanied by four fermionic supersymmetric partners denoted schematically as $\Psi$.
$[d^{4P-4}\mu_{B}^{\rm Rel.}]$ in ${\mathcal Z}^{(P)}$ corresponds to the remaining bosonic zero mode integration measure 
over the $(4P-4)$-dimensional relative monopole moduli space, and $[d^{4P-4}\mu_F^{\rm Rel.}]$ is their fermionic counterparts.

\paragraph{}

The evaluation of the centre of mass contribution $\mathcal{G}^{(4)}_{\rm COM}$ is, in fact, almost identical to the one for single monopole correlation function, i.e., $P=1$, with the essential modification to the one-loop ratio of determinants $\mathcal{R}^{(P)}$, which will be discussed below. 
As far as the zero mode integration measure is concerned, we can adapt the results in \cite{CDP}:
\begin{equation}
\begin{aligned}
&\int [d^{3}X(x^{4})] [d\Phi(x^{4})] [d^{4}\Psi(x^{4})]
\exp\left(-\int^{2\pi R}_{0} L_{\rm COM}\right)\\
&=\int [d^{3}X] [d^{4}\Psi]
\left(2\pi \sqrt{\frac{R}{M_P}}\right)
\sum_{N_{e} \in {\mathbb Z}}\exp\left(-\pi R \, \frac{|a_P|^{2}}{M_P}N_{e}^{2}\right)
\end{aligned}
\end{equation} 
where $N_{e}$ is the overall global $U(1)$ electric charge, $M_P$ is $P$-monopole mass, and $a_P = \sum_{B=1}^{N-1} (\vec a\vec\alpha_{B})$.

\paragraph{}

The contribution from the relative monopole moduli space $\mathcal{Z}^{(P)}$ is more interesting, as it is this part which dictates the change in the BPS index $\Omega(\gamma, \vec a)$.
At this point, we can recall from \cite{Dorey 2000, DHK} that the four-fermion correlation $\mathcal{G}_{4}^{(P)}$ in the compactified gauge theory on ${\mathbb R}^{3}\times S^{1}$ can, in fact, be regarded as a refinement of the Witten index, which traces over the BPS states, that is,
\begin{equation}
\label{hamiltonian}
\mathcal{G}_{4}(y_{1},y_{2},y_{3},y_4)={\rm Tr}_{\rm BPS}
\left(
(-1)^F \prod^{2}_{A=1}\rho_{1}(y_{2A-1})\rho_{2}(y_{2A}) \exp(-2\pi R \, H_{\rm QM})
\right)
\end{equation}
where $H_{\rm QM}$ is the hamiltonian associated with the supersymmetric quantum mechanic lagrangian $L_{\rm QM}$ over the monopole moduli space.
This, in particular, allows us to relate $\mathcal{Z}^{(P)}$ to the index-like computation over the $P$-monopole relative moduli space:
\begin{equation}
\label{ZPindex}
\mathcal{Z}^{(P)}={\rm Tr}_{(P)}\left( (-1)^F \exp\left( -2\pi R \, H_{\rm Rel.}^{(P)} \right) \right)
\,.
\end{equation}
To be more precise, as we are keeping the compactfied radius $R$ fixed and arbitrary for the time being, 
the quantitiy $\mathcal{Z}^{(P)}$ corresponds to the so-called ``bulk'' contribution to the usual $L^{2}$-normalisable index $\mathcal{I}_{L^{2}}$ over the monopole moduli space \cite{DHK}.

\paragraph{}

The most interesting case is the one with $P=2$, i.e., with two distinct simple monopoles, other $P>2$ cases can be discussed analogously. 
The relative moduli space for two distinct monopoles with equal masses 
is known to be Taub--NUT space \cite{LWY}, in the presence of general complex scalar VEV, 
the relative electric charge between them induces additional potential~\footnote{
after promoting each one of them into simple dyon through semiclassical quantisation
}, which is proportional to the square norm of a tri-holomorphic vector field $G_m$ over the Taub--NUT space.
The problem of computing the index $\mathcal{Z}^{(2)}$ (\ref{hamiltonian}) or number of bound states/composite monopole can be mapped to counting the number of normalisable solutions of the Dirac equation in such background (Taub--NUT $+$ potential) \cite{GKPY, Stern Yi}:
\begin{equation}
\label{DiracEqn}
-\gamma^m \cdot (i\nabla_m + G_m)\Psi=0
\end{equation}
where $\nabla_m$ is the covariant derivative over Taub--NUT space, $\gamma^m$ is the gamma matirx.
This counting problem has been solved in \cite{Pope}, and the result is that for two simple dyons of charges $(n_e^{1} \vec\alpha_{1}, \vec\alpha_{1})$
and $(n_e^{2} \vec\alpha_{2}, \vec\alpha_{2})$, there are $|n_e^1-n_2^2|=2|n_e^{-}|$ solutions of (\ref{DiracEqn}) existing if the inequality 
$|n_e^1-n_e^2|< 16 \pi g^{-4} |\sin \Delta \phi|$ where $2 \Delta \phi=\phi_{W_1}-\phi_{W_2}$ is satisfied.
We see that at the generic point in the weakly coupled region of the moduli space, $1/g_{\rm eff}^{2} \gg 1$, this inequality can be easily satified, however, at the wall of marginal stability, $\Delta \phi=0$, and the bound states disappear, as expected. 
This precisely matches with the discontinuous change in the BPS index $\Delta \Omega(\gamma, \vec a)$, and our remaining task would be to explain how the crucial one-loop factor can arise from semiclassical computations.


\subsection{Semiclassical derivation of one-loop determinants}

\paragraph{}

In \cite{CDP}, it was shown that the one-loop factor from $W$ bosons with electric charges $\pm\vec\alpha_{A}$, i.e., $\mathcal{D}_{\gamma_{A},A}(-e^{i\phi_{\gamma_{A}}})$ in (\ref{D factor simple}), can be derived directly by considering the non-zero mode fluctuations around the associated $SU(2)$ monopole.
To see how additional contributions $\mathcal{D}_{\gamma_{A},B\ne A}(-e^{i\phi_{\gamma_{A}}})$ in (\ref{D factor mixed}, \ref{D factor kernel}) can also be obtained from semiclassical analysis, the key is to adapt the difference of the densities of states $\delta\rho_{A}(x^{2})$ in the pure $SU(2)$ theory to the $SU(n)$ case.
We can work this out by considering the index function $\mathcal{I}(\mu^{2})=\sum_{B} {\mathcal{I}}_{B}(\mu^{2})$ counting the zero modes in the context of three-dimensional instanton computation for higher-rank gauge groups \cite{Fraser Tong}.
For completeness, we first write down the index function for the zero mode fluctuations charged under the same root $\vec\alpha_{A}$:
\begin{equation}
\label{IA}
\mathcal{I}_{A}(\mu^{2}) =
\frac{2 M_{W_{A}}}{(M_{W_{A}}^{2}+\mu^{2})^{1/2}}
\end{equation}
where $M_{W_{A}}=|Z_{W_{A}}|$ is the mass of the $W$ boson charged under $\vec\alpha_{A}$.
For the fluctuations charged under $\vec\alpha_{B\ne A}$, simple manipulation gives the index function (see equation (15) in \cite{Fraser Tong}):
\begin{equation}
\label{FTdensity}
\mathcal{I}_{B}(\mu^{2}) =
\frac{2(\vec\alpha_{A}\cdot\vec\alpha_{B})M_{W_{B}}}{\left( M_{W_{B}}^{2}+\mu^{2} \right)^{1/2}}
\frac{(\lambda_{A}^{i}\lambda_{B}^{i}) \, \mu^{2}}{\left( \mu^{2}+{M_{W_{B}}^{2}}(1-(\lambda_{A}^{i}\lambda_{B}^{i})^{2}) \right)}
\end{equation}
where $M_{W_{B}}=|Z_{W_{B}}|$ is the mass of the $W$ boson charged under $\vec\alpha_{B}$, $\lambda_{B}^{i}=(\vec v\,^{i}\vec\alpha_{B})/||\vec v\,^{l}\vec\alpha_{B}||_{l}$ is the three-dimensional analogue of phase angle where $\vec{v}\,^{i}=(\re\vec a,\im\vec a,\vec\theta_{e}/2\pi R)^{i}$ is an $SO(3)$ three-vector (with respect to superscript $i$) consisting of the three adjoint scalars belonging to the three-dimensional vector multiplet.
We can now use the identity for the index function used in \cite{Kaul} to derive the difference in the density of states in our case:
\begin{equation}
\label{DiffI}
\mathcal{I}_{B}(\mu^{2})-\mathcal{I}_{B}(0) = \int^{\infty}_{0}dx^{2} \frac{\mu^{2}}{x^{2}+\mu^{2}}\delta \rho_{B}(x^{2})\,.
\end{equation}
Using our earlier results for gauge group $SU(2)$, we can derive the required $\delta\rho_{B}(x^{2})$:
\begin{equation}
\begin{aligned}
\delta\rho_{B}(x^{2}) = & -\frac {2(\vec\alpha_{A}\cdot\vec\alpha_{B})M_{W_{B}}}{\pi}\frac{\Theta(x^{2}-M_{W_{B}}^{2})}{x^{2}(x^{2}-M_{W_{B}}^{2})^{1/2}}
\frac{x^{2}(\lambda_{A}^i\lambda_{B}^i)}{x^{2}-M_{W_{B}}^{2}(\lambda_{A}^i\lambda_{B}^i)^{2}}
\\
& +2\delta(x^{2}-M_{W_{B}}^{2}(1-(\lambda_{A}^i\lambda_{B}^i)^{2}))
\end{aligned}
\end{equation}
where $\Theta(y)$ is a step function such that $\Theta(y)=1$ if $y\geq 0$ and $\Theta(y)=0$ if $y<0$.
We can now set $x=M_{W_{B}}\cosh t$ and rearrange $dx^{2}\delta\rho_{B}(x^{2})$ into
\begin{equation}
\label{DOS2}
\int^{\infty}_{0} dx^{2} \delta\rho_{B}(x^{2})=-\frac{4(\vec\alpha_{A}\cdot\vec\alpha_{B})}{\pi}\int^{\infty}_{0} dt \, \frac{\cosh t \, (\lambda_{A}\lambda_{B})}{\cosh^{2} t-(1-(\lambda_{A} \lambda_{B})^{2})}
\,.
\end{equation}
By using the $SO(3)$ symmetry to rotate into the vacuum $\vec\theta_{e}=0$, the difference in the densities of states $\delta \rho_{B}(x^{2})$ obtained here for ${\mathbb R}^{3}$ can be identified with the corresponding quantities for ${\mathbb R}^{3}\times S^{1}$. 
From the definition of $\lambda_{A}^{i}$, it follows that
\begin{equation}
\lambda_{A}^{i}\lambda_{B}^{i} = \cos\Delta\phi
\,,
\end{equation}
where $\Delta\phi$ was introduced in (\ref{D factor kernel}), and we see that $dx^{2}\delta\rho_{B}(x^{2})$ can be identified with $dt\rho(t,\Delta\phi)$ given in (\ref{D factor mixed}, \ref{D factor kernel}) up to an overall numerical factor.
At this point, we can repeat the analysis in \cite{CDP}, where enumeration of non-zero mode fluctuations in the monopole background in ${\mathbb R}^{3}\times S^{1}$ was mapped to the partition function of harmonic oscillators with inverse temperature $2\pi R$ and background chemical potential $\theta_{e}/2\pi R$.
This yields the additional logarithmic integrands appearing in (\ref{D factor mixed}).
The overall factor can be fixed by requiring that for $B=A$, the formula reproduces the $SU(2)$ one-loop factor.
This completes our semiclassical derivation of the additional one-loop factor $\mathcal{D}_{\gamma_{A},B}(-e^{i\phi_{\gamma_{A}}})$.

\paragraph{}

We can also consider the semiclassical one-loop determinant in the strict three-dimensional limit:
\begin{equation}
\label{3Dlimit}
2\pi R\to 0
\,,
\quad
\left( \re\vec a,\ \im\vec a,\ \frac{\vec\theta_{e}}{2\pi R} \right) = \const
\,.
\end{equation}
The semiclassical one-loop factor in this case reduces to \cite{DKMTV}
\begin{equation}
\mathcal{R}^{\rm (3D)}=\lim_{\kappa\to 0}\left( \kappa^{2}\exp\left(\int^{\infty}_{\kappa}\frac{d\nu}{\nu}{\mathcal I}(\nu)\right) \right)^{1/2}
\,.
\end{equation}
If we substitute the index function $\mathcal{I}_{B}(\mu^{2})$ (\ref{FTdensity}) into this expression and exchange the order of $x^{2}$ and $\nu$ integrations, we obtain:
\begin{equation}
\label{logR3D}
\begin{aligned}
\log {\mathcal{R}}^{\rm (3D)}
& = \lim_{\kappa \to 0}
\left(
\log \kappa+\frac{1}{2}\sum_{B}
\left(
\int^{\infty}_{0}dx^{2}\delta \rho_{B}(x^{2})
\left( \log(\nu+x^{2}) \right)^{\infty}_{\kappa} + \mathcal{I}_{B}(0)\left( \log\nu\right )^{\infty}_{\kappa}
\right)
\right)
\\
& = -\frac{1}{2}\sum_{B}\int^{\infty}_{0}dx^{2}\delta \rho_{B}(x^{2})\log(x^{2}) + ({\rm cutoffs})
\,.
\end{aligned}
\end{equation}
The same result can be obtained by considering the one-loop factor $\mathcal{D}_{\gamma_{A},B}(-e^{i\phi_{\gamma_{A}}})$ given in (\ref{D factor mixed}): in the three-dimensional limit (\ref{3Dlimit}), the logarithmic integrands in $\mathcal{D}_{\gamma_{A},B}(-e^{i\phi_{\gamma_{A}}})$ become
\begin{equation}
\begin{aligned}
& \log\left( 1-e^{-2\pi R |Z_{W_{B}}|\cosh t+i\theta_{W_{B}}} \right)+\log\left( 1-e^{-2\pi R |Z_{W_{B}}|\cosh t-i\theta_{W_{B}}} \right)
\\
& \to \log\left( |Z_{W_{B}}|^{2}\cosh^{2}t+\left( \frac{\theta_{W_{B}}}{2\pi R} \right)^{2} \right)+2\log(2\pi R)
\,,
\end{aligned}
\end{equation}
then, after substituting $x=M_{W_{B}}\cosh t$, rotating into the vacuum where $\vec\theta_{e}/2\pi R=\vec 0$, and combining with the earlier identification of the density of states, we can see that in the limit $R\to 0$, $\mathcal{D}_{\gamma_{A},B}(-e^{i\phi_{\gamma_{A}}})$ corresponds to the ratio of determinants (\ref{logR3D}). 

\paragraph{}

We have calculated the ratio of one-loop determinants in $\mathbb{R}^{3}\times S^{1}$ and matched it with the GMN prediction extracted in the previous section (\ref{D factor 3}, \ref{D factor simple}, \ref{D factor mixed}, \ref{D factor kernel}).
Finding the one-instanton action and the overall coefficient for the moduli space metric is essentially equivalent to the $SU(2)$ case \cite{CDP}.
Summing up, we conclude that the one-instanton metric calculated semiclassically coincides with the prediction (\ref{metric NP1}) obtained in the previous section.


\subsection{Interpolating to three dimensions}

\paragraph{}

In this section, we shall start from the semiclassical expansion of the moduli space metric on $\mathbb{R}^{3}\times S^{1}$ and demonstrate how the smoothness of the moduli space metric persists in three dimensions.
In \cite{Fraser Tong}, it was shown how the corresponding three-dimensional metric remains smooth as the VEV crosses the wall of marginal stability; in the following, we shall discuss how our results can be related to the ones there and demonstrate that our one-instanton correction in three dimensions coincides with the one obtained in \cite{Fraser Tong}.

\paragraph{}

We shall focus on the mixing terms between dyons of charges $\gamma_{1}=(n_{e}^{1}\vec\alpha_{1},\vec\alpha_{1})$ and $\gamma_{2}=(n_{e}^{2}\vec\alpha_{2},\vec\alpha_{2})$ in (\ref{symplectic form NP2}), where $\vec\alpha_{1,2}$ are the two simple roots, $\{n_{e}^{1},n_{e}^{2}\}\subset\mathbb{Z}$.
After Poisson-resumming over suitable combination of their electric charges, we shall demonstrate that these contributions again combine to give  one-monopole correction of magnetic charge $\vec\alpha_{1}+\vec\alpha_{2}$ near the wall of marginal stability in three dimensions.
A similar computation for the one-instanton term has been performed in \cite{CDP}, however, an important difference here is that we perform the Poisson resummation directly before integrating over the spectral parameters: this preserves the integration kernel, which is crucial for ensuring the smoothness.

\paragraph{}

At leading order in $g_{\rm eff}^{2}$ expansion, the relevant terms from (\ref{symplectic form NP2}) for our analysis can be shown to be
\begin{equation}
\label{Poiwnp2}
\begin{aligned}
\frac{R}{(2\pi i)^{2}}\sum_{n_{e}^{\pm}\in{\mathbb Z}} & \Omega(\gamma_{1},\vec a)\Omega(\gamma_{2},\vec a)\sigma(\gamma_{1})\sigma(\gamma_{2})n_{e}^{-}
\int^{\infty}_{0} \frac{dy'}{y'}\int^{\infty}_{0} \frac{dy''}{y''}
\\
& \mathcal{X}^{(0)}_{\gamma_{1}}\left( -y'e^{i\phi_{1}} \right)\mathcal{X}^{(0)}_{\gamma_{2}}\left( -y''e^{i\phi_{2}} \right)
\left( \frac{y' e^{i\phi_{1}}+y'' e^{i\phi_{2}}}{y'e^{i\phi_{1}}-y'' e^{i\phi_{2}}} \right)
\\
& \left(
dZ_{\gamma_{1}}\wedge d\bar{Z}_{\gamma_{1}}+\left(\frac{y'' e^{i\phi_{2}}}{y'e^{i\phi_{1}}+\frac{1}{y'e^{i\phi_{1}}}}
+\frac{\frac{1}{y'e^{i\phi_{1}}}}{y'e^{i\phi_{2}}+\frac{1}{y'e^{i\phi_{2}}}}\right)dZ_{\gamma_{1}}\wedge d\bar{Z}_{\gamma_{2}} 
+ (1\leftrightarrow 2~;~ y'\leftrightarrow y'' )
\right)
\end{aligned}
\end{equation}
where we have introduced $n_{e}^{\pm}=\frac{1}{2}(n_{e}^{1}\pm n_{e}^{2})$.
To perform Poisson resummation before the $y'$ and $y''$ integrations, we first expand the expressions in exponents coming from the Darboux coordinates:
\begin{equation}
\label{2instexp}
\begin{aligned}
& -\pi R |Z_{\gamma_{1}}|\left( y'+\frac{1}{y'} \right)-\pi R |Z_{\gamma_{2}}|\left( y''+\frac{1}{y''} \right) + i\theta_{\gamma_{1}}+i\theta_{\gamma_{2}} \simeq
\\
& -\frac{4\pi^{2}R}{g_{\rm eff}^{2}}\left(|\vec a \vec\alpha_{1}|\left( y'+\frac{1}{y'} \right)+|\vec a \vec\alpha_{2}|\left( y''+\frac{1}{y''} \right)\right)+ i\vec\theta_{m}(\vec\alpha_{1}+\vec\alpha_{2})
\\
& -\frac{\pi R}{2} M(y',y'')\left(n_{e}^{+} + n_{e}^{-}\frac{m_{1}\left( y'+\frac{1}{y'} \right)-m_{2}\left( y''+\frac{1}{y''} \right)}{m_{1}\left( y'+\frac{1}{y'} \right)+m_{2}\left( y''+\frac{1}{y''} \right)}\right)^{2}+in_{e}^{+}\vec\theta_{e} (\vec\alpha_{1}+\vec\alpha_{2})\\
& -\frac{\pi R}{2} m(y',y'')(2n_{e}^{-})^{2} +i n_{e}^{-}\vec\theta_{e} (\vec\alpha_{1}-\vec\alpha_{2})
\end{aligned}
\end{equation}
where, without loss of generalities, we have set the four-dimensional topological angle $\Theta_{\rm eff}=0$ and defined the following quantities:
\begin{align}
& M(y',y'')=m_{1}\left( y'+\frac{1}{y'} \right)+m_{2}\left( y''+\frac{1}{y''} \right)
\,,
\quad
\frac{1}{m(y',y'')}=\frac{1}{m_{1}\left( y'+\frac{1}{y'} \right)}+\frac{1}{m_{2}\left( y''+\frac{1}{y''} \right)}
\,,
\\
& m_{1,2}=\frac{g_{\rm eff}^{2}}{4\pi}|\vec a \vec\alpha_{1,2}|
\,.
\end{align}
We shall sum over the overall electric charge $n_{e}^{+}$, while keeping fixed the relative electric charge $n_{e}^{-}$, and at the order of our $g_{\rm eff}^{2}$ expansion, we only need to sum over the terms in (\ref{2instexp}):
\begin{equation}
\label{Poisson2inst}
\begin{aligned}
\sum_{n_{e}^{+}\in {\mathbb Z}}\exp\left(
-\frac{\pi R}{2} M(y',y'')\left(n_{e}^{+} + n_{e}^{-}\frac{m_{1}\left( y'+\frac{1}{y'} \right)-m_{2}\left( y''+\frac{1}{y''} \right)}{m_{1}\left( y'+\frac{1}{y'} \right)+m_{2}\left( y''+\frac{1}{y''} \right)}\right)^{2}+in_{e}^{+}\vec\theta_{e} (\vec\alpha_{1}+\vec\alpha_{2})
\right)
\\ 
=
\sum_{k\in {\mathbb Z}}\sqrt{\frac{8}{RM(y',y'')}}\exp\left(
-\frac{2 \omega_k^{2}}{\pi RM(y',y'')}
-i\omega_k n_{e}^{-}\frac{m_{1}\left( y'+\frac{1}{y'} \right)-m_{2}\left( y''+\frac{1}{y''} \right)}{m_{1}\left( y'+\frac{1}{y'} \right)+m_{2}\left( y''+\frac{1}{y''} \right)}
\right)
\end{aligned}
\end{equation}
where $\omega_k=\vec{\theta}_{e}(\vec\alpha_{1}+\vec\alpha_{2})/2-2\pi k$.
In the three-dimensional limit, only $k=0$ term in the summation above survives, all other $k\neq 0$ terms are suppressed, furthermore, all terms depending on $n_{e}^{-}$ in (\ref{2instexp}) and (\ref{Poisson2inst}) vanish in such limit.
After further rotating into the $\vec\theta_{e}=0$ vacuum, the integrand in (\ref{Poiwnp2}) in the three-dimensional limit becomes
\begin{equation}
\begin{aligned}
\label{3Dmetric}
\frac{1}{(2\pi i)^{2}}\left(\frac{2\pi}{e_{\rm eff}^{2}}\right)^{5/2} & \int^{\infty}_{0} \frac{dy'}{y'}\int^{\infty}_{0} \frac{dy''}{y''}\left(\frac{y' e^{i\phi_{1}}+y'' e^{i\phi_{2}}}{y'e^{i\phi_{1}}-y'' e^{i\phi_{2}}}\right) \\ &
\frac{\mathcal{D}_{\gamma_{1}}\left( -y'e^{i\phi_{1}} \right)\mathcal{D}_{\gamma_{2}}\left( -y''e^{i\phi_{2}} \right)}{(2\pi R)^{2}\left( |\vec a\vec\alpha_{1}|\left( y'+\frac{1}{y'} \right)+|\vec a\vec\alpha_{2}|\left( y''+\frac{1}{y''} \right) \right)^{1/2}}
\\
& \exp\left( -\frac{2\pi}{e_{\rm eff}^{2}} \left( |\vec a\vec\alpha_{1}|\left( y'+\frac{1}{y'} \right)+|\vec a\vec \alpha_{2}|\left( y''+\frac{1}{y''} \right) \right)+i\vec\theta_{m}(\vec\alpha_{1}+\vec\alpha_{2}) \right)
\,.
\end{aligned}
\end{equation}
Near the wall of marginal stability, as in the finite radius situation, we can again evaluate the $y'$ or $y''$ integration using the Cauchy residue theorem, and after setting $y'=y''$ and $\phi_{\gamma_{1}}=\phi_{\gamma_{2}}$, we see that (\ref{3Dmetric}) indeed goes over to the metric correction corresponding to a monopole charged under the composite root $\vec\alpha_{1}+\vec\alpha_{2}$.
It was also noted in \cite{Fraser Tong} that there are additional singularities in the one-loop factor as we approach the walls of marginal stability; these singularities are cancelled by the fact that the associated index $\Omega(\gamma,\vec a)$ also changes discontinuously to zero there, hence, the overall moduli space metric remains smooth.
This also echoes the ``soft modes'' computations done in \cite{Fraser Tong}, which is nothing but the zero radius limit of the moduli space quantum mechanics described in section \ref{sec: instanton correlator} for $\Omega(\gamma,\vec a)$.  
We have thus demonstrated that the smoothness property of the moduli space metric persists in the appropriate three-dimensional limit, as should be expected.

\paragraph{}

The Poisson resummation also allows us to interpolate to the one-instanton correction to the moduli space metric (\ref{metric NP1}) in three dimensions \cite{Fraser Tong}.
For each positive root $\vec\alpha_{A}$, we can Poisson-resum all terms corresponding to dyons with magnetic charge $\vec\alpha_{A}$ (terms corresponding to $-\vec\alpha_{A}$ are their complex conjugates).
Again, we split the relevant one-loop factor, $\mathcal{D}_{(\vec 0,\vec\alpha_{A})}(-ie^{i\phi_{W_{A}}})$, into the $A$ term and $B\ne A$ terms (\ref{D factor 3}).
The $A$ term in this limit is known to be $\mathcal{D}_{(\vec 0,\vec\alpha_{A}),A}(-ie^{i\phi_{W_{A}}})=(4\pi RM_{W_{A}})^{2}$ \cite{CDP}, the three-dimensional values of $B\ne A$ terms were calculated in the previous section.
After Poisson-resumming and taking the limit $R\to 0$, we find the following result:
\begin{equation}
g_{a^{I}\bar a^{J},A} =
\frac{16\pi}{e_{\rm eff}^{4}}
M_{W_{A}} \left( \prod_{B\ne A}\mathcal{D}_{(\vec 0,\vec\alpha_{A}),B}(-ie^{i\phi_{W_{A}}}) \right)
\exp\left( -\frac{4\pi}{e_{\rm eff}^{2}}M_{W_{A}}+i\vec\alpha_{A}\vec\theta_{m} \right)
\end{equation}
where $1/e_{\rm eff}^{2}=2\pi R/g_{\rm eff}^{2}$ is the effective gauge coupling in three dimensions.
Using this metric, we can compute the Riemann tensor and then recover the coefficient of the four-fermion correlation function (or, more precisely, the bosonic partner) \cite{Fraser Tong}.

\subsection*{Acknowledgements}

HYC is generously supported in part by NSF CAREER Award No. PHY-0348093, DOE grant DE-FG-02-95ER40896, a Research Innovation Award and a Cottrell Scholar Award from Research Corporation, and a Vilas Associate Award from the University of Wisconsin.
KP is supported by a research studentship from Trinity College, Cambridge.

\appendix


\section{Pentagon wall-crossing formulae for composite dyons}
\label{sec: pentagon}

\paragraph{}

First, let us show how to change the basis of charge-vectors in a wall-crossing formula.
Any given formula
\begin{equation}
\prod_{k=1}^{K} \mathcal{K}_{\gamma_{k}}=1
\,,
\
K\in\mathbb{N}\cup\{+\infty\}
\,,
\end{equation}
can be re-expressd in different coordinates ($\gamma_{k}\to\beta_{k}$) if the transformation of charge-vectors is linear and if their symplectic product remains the same for any pair of charges in the formula, i.e., $\langle\beta_{i},\beta_{j}\rangle=\langle\gamma_{i},\gamma_{j}\rangle$.
The formula in these new coordinates is
\begin{equation}
\prod_{k=1}^{K}\mathcal{K}_{\beta_{k}}=1
\,.
\end{equation}

\paragraph{}

Let us prove this statement. Suppose that we change coordinates as $\gamma_{(i)}\to\beta_{(i)}$ for all possible charges $\gamma_{(i)}$.
Linearity of the transformation ensures that all symplectic products are also linear, i.e., $\langle\beta_{(1)}+\beta_{(2)},\beta_{(3)}\rangle=\langle\beta_{(1)},\beta_{(3)}\rangle+\langle\beta_{(2)},\beta_{(3)}\rangle$, and that changing the coordinates does not violate the condition that $\mathcal{K}_{\beta_{(1)}+ \beta_{(2)}}=\mathcal{K}_{\beta_{(1)}}\mathcal{K}_{\beta_{(2)}}$.
The operators $\mathcal{K}_{\beta_{k}}$ act depending only on the symplectic products between $\beta_{k}$ and $\beta_{l}$ where $k<l\le K$, which are conserved.

\paragraph{}

We consider the standard pentagon wall-crossing formula \cite{KS}:
\begin{equation}
\mathcal{K}_{(\frac{1}{2},0)}\mathcal{K}_{(0,1)} =
\mathcal{K}_{(0,1)}\mathcal{K}_{(\frac{1}{2},1)}\mathcal{K}_{(\frac{1}{2},0)}
\,,
\quad
\mathcal{K}_{(0,1)}\mathcal{K}_{(\frac{1}{2},0)} =
\mathcal{K}_{(\frac{1}{2},0)}\mathcal{K}_{(\frac{1}{2},1)}\mathcal{K}_{(0,1)}
\,.
\end{equation}
To prove it, one just needs to check the equality on a basis of Darboux coordinates.
In the case of $r$ electric and $r$ magnetic charges, the equations are
\begin{equation}
\begin{aligned}
\mathcal{K}_{((0,0,\dots,0),(1,0,\dots,0))}\mathcal{K}_{((\frac{1}{2},0,\dots,0),(0,0,\dots,0))} =
\mathcal{K}_{((\frac{1}{2},0,\dots,0),(0,0,\dots,0))}\mathcal{K}_{((\frac{1}{2},0,\dots,0),(1,0,\dots,0))}\mathcal{K}_{((0,0,\dots,0),(1,0,\dots,0))}
\,,
\\
\mathcal{K}_{((\frac{1}{2},0,\dots,0),(0,0,\dots,0))}\mathcal{K}_{((0,0,\dots,0),(1,0,\dots,0))} =
\mathcal{K}_{((0,0,\dots,0),(1,0,\dots,0))}\mathcal{K}_{((\frac{1}{2},0,\dots,0),(1,0,\dots,0))}\mathcal{K}_{((\frac{1}{2},0,\dots,0),(0,0,\dots,0))}
\,.
\end{aligned}
\end{equation}
It is easy to see that this is correct: the relation is known to be valid when the left and the right-hand sides act on $\mathcal{X}_{e}^{1}$ and $\mathcal{X}_{m\,1}$; both sides give identity when acting on $\mathcal{X}_{e}^{I}$ and $\mathcal{X}_{m\,I}$ for $I>1$.
Therefore, more generally, changing the basis, the pentagon equation for any $r$ is (\ref{pentagon}).

\paragraph{}

The formula can be applied to the decay process of the composite dyons.
Indeed, in (\ref{SU(3) composite dyon decay}), the symplectic product of the two simple dyons is
\begin{equation}
\left\langle
\pm (p\vec\alpha_{1},\vec\alpha_{1})
\,,
\pm ((p+1)\vec\alpha_{2},\vec\alpha_{2})
\right\rangle
= -\frac{1}{2}
\,,
\end{equation}
and we obtain (\ref{SU(3) pentagon}).
The same applies to the $SU(n)$ composite dyons: after some algebra, we see that in (\ref{composite dyon decay}),
\begin{equation}
\begin{aligned}
& \quad \left\langle
\pm \left(
p\sum_{m=i}^{k}\vec\alpha_{m}+\sum_{l=i+1}^{k}\epsilon_{l}\sum_{m=l}^{k}\vec\alpha_{m}
\,, \
\sum_{m=i}^{k}\vec\alpha_{m}
\right)
\,,
\right.
\\
& \quad \quad \left.
\pm \left(
\left( p+\sum_{l=i+1}^{k}\epsilon_{l} \right)\sum_{m=k+1}^{j-1}\vec\alpha_{m}+\sum_{l=k+1}^{j-1}\epsilon_{l}\sum_{m=l}^{j-1}\vec\alpha_{m}
\,, \
\sum_{m=k+1}^{j-1}\vec\alpha_{m}
\right)
\right\rangle
\\
= & \left\langle
\left(
\left( p+\sum_{l=i+1}^{k}\epsilon_{l} \right)\vec\alpha_{k}
\,, \
\vec\alpha_{k}
\right)
\,,
\left(
\left( p+\sum_{l=i+1}^{k}\epsilon_{l} \right)\vec\alpha_{k+1}+\epsilon_{k+1}\vec\alpha_{k+1}
\,, \
\vec\alpha_{k+1}
\right)
\right\rangle
= -\frac{\epsilon_{k+1}}{2} = \pm\frac{1}{2}
\,,
\end{aligned}
\end{equation}
and we obtain the general pentagon formula (\ref{SU(n) pentagon}).


\end{document}